\documentclass[journal]{IEEEtran}
\ifCLASSINFOpdf
\else
\fi
\usepackage{theorem}
\theorembodyfont{\rmfamily}

\usepackage{amsmath,amssymb}
\newtheorem{assumption}{Assumption}
\newtheorem{proposition}{Proposition}
\newtheorem{lemma}{Lemma}
\newtheorem{theorem}{Theorem}
\newtheorem{definition}{Definition}
\newtheorem{problem}{Problem}

\newtheorem{remark}{Remark}

\renewcommand{\IEEEQED}{{\hfill \IEEEQEDopen}\\}

\usepackage{graphicx,color}

\usepackage{cite}

\newcommand{\bm}[1]{\boldsymbol{#1}}

\DeclareMathOperator*{\sfmax}{\mathsf{max}}

\DeclareMathOperator{\sfsin}{\mathsf{sin}}
\DeclareMathOperator{\sfcos}{\mathsf{cos}}

\DeclareMathOperator{\sfdiag}{\mathsf{diag}}
\DeclareMathOperator{\sfcol}{\mathsf{col}}
\DeclareMathOperator{\sfdet}{\mathsf{det}}

\newcommand{\mat}[1]{\left[\: \begin{matrix} #1 \end{matrix} \:\right]}
\newcommand{\spliteq}[1]{\begin{split} #1 \end{split}}
\newcommand{\simode}[1]{\left\{\:  \begin{split} #1 \end{split} \right.}

\begin{document}
%
\title{
Modularity-in-Design of Dynamical Network Systems: Retrofit Control Approach
}
%
%
%

\author{
        Takayuki~Ishizaki,~\IEEEmembership{Member,~IEEE,}
        Hampei~Sasahara,~\IEEEmembership{Member,~IEEE,}
        Masaki~Inoue,~\IEEEmembership{Member,~IEEE,}
        Takahiro~Kawaguchi,~\IEEEmembership{Member,~IEEE,}
        and~Jun-ichi~Imura,~\IEEEmembership{Senior Member,~IEEE}
\thanks{T.~Ishizaki, T.~Kawaguchi, and J.~Imura are with Graduate School
of Engineering, Tokyo Institute of Technology, Meguro,
Tokyo, 152-8552 Japan. e-mail: \{ishizaki, kawaguchi, imura\}@cyb.sc.e.titech.ac.jp.}
\thanks{H.~Sasahara is with School of Electrical Engineering and Computer Science, KTH Royal Institute of Technology;
SE-100 44, Stockholm, Sweden.
 e-mail: hampei@kth.se.}
\thanks{M.~Inoue is with the Faculty of Science and Technology, Keio University, Yokohama, Kanagawa, 223-8551 Japan. e-mail: minoue@appi.keio.ac.jp.}
\thanks{Manuscript received xxxx; revised xxxx.}}

%
%

\markboth{Journal of \LaTeX\ Class Files,~Vol.~X, No.~X, XXXX}%
{Shell \MakeLowercase{\textit{et al.}}: Bare Demo of IEEEtran.cls for IEEE Journals}
%



\maketitle

\begin{abstract}
In this paper, we develop a modular design method of decentralized controllers for linear dynamical network systems, where multiple subcontroller designers aim at individually regulating their local control performance with accessibility only to their respective subsystem models.
First, we derive a constrained version of the Youla parameterization that characterizes all retrofit controllers for a single subcontroller, defined as an add-on type subcontroller that manages a subsystem.
The resultant feedback system is kept robustly stable for any variation in the neighboring subsystems, other than the subsystem of interest, provided that the original system is stable prior to implementing the retrofit control. 
Subsequently, we find out a unique internal structure of the retrofit controllers, assuming that the interaction input signal from the neighboring subsystems is measurable.
Furthermore, we show that the simultaneous implementation of multiple retrofit controllers, designed by individual subcontroller designers, can improve the upper bound of the overall control performance. 
Finally, the practical significance of the method is demonstrated via an illustrative example of frequency regulation using the IEEE 68-bus power system model.
\end{abstract}


%
\IEEEpeerreviewmaketitle

\section{Introduction}
%
%
%
%

\IEEEPARstart{M}{odular} design or modularity-in-design is a widely accepted concept of system design for managing the complexity of large-scale system designs, to enable parallel work via multiple independent entities or developers, and to make future modifications in the subsystems or modules flexible.
In the field of software engineering, the necessity and benefits of modularization have been presented in the seminal paper \cite{parnas1972criteria}, where a module was introduced as a distinct unit of work that can be managed by one developer without considerable efforts for the adjustment or coordination with other developers.
Thereafter, the notion of modular design has been significantly expanded, and its advantages have been analyzed in a number of studies \cite{huang1998modularity,schilling2000toward,baldwin2000design,baldwin2006modularity}.
Notably, \cite{baldwin2000design} and \cite{baldwin2006modularity} reported that the modularization of products can induce and facilitate the modularization of industrial structures, as recently exemplified by the computer industry.
A modular design approach is often compared to a contrastive approach called integral design \cite{ulrich2003role,ulrich2003product}, where a single authority or designer performs high-level integration of interdependent components.

In the research on systems and control engineering, relevant problems of analyzing and synthesizing large-scale network systems have been discussed for several decades.
In particular, a wide variety of decentralized and distributed control methods have been devised from different perspectives; see, e.g.,\cite{siljak1972stability,rotkowitz2006characterization,langbort2004distributed,vsiljak2005control} and references therein.
However, a majority of the existing methods are not classified as modular design approaches; they are classified as integral design approaches of structured controllers, where a single authority uses an entire network model to simultaneously design all subcontrollers of a decentralized or distributed controller.
Therefore, a small change in a single subsystem or subcontroller may require a significant change in all other components.

On the other hand, several control methods that can be classified as modular design approaches have been reported by previous studies.
To the best of the authors' knowledge, the work \cite{langbort2010distributed,delvenne2006price} is the first to formally discuss a control problem where subcontrollers are designed in a modular fashion.
In this seminal work, the performance limitation of linear quadratic regulators is analyzed using the notions referred to as the competitive ratio and domination metrics, with the assumption that each subsystem is one-dimensional and has its own input port.
This analysis is further generalized to the case of multi-dimensional subsystems with the assumption that the state of each subsystem can be fully actuated \cite{farokhi2013optimal}.

As a different approach, \cite{petes2017scalable} developed a scalable design method of stable networks, based on the framework of integral quadratic constraints (IQCs).
In particular, a general stability criterion that can be tested in a decentralized manner is derived by introducing a subsystem structure in the IQC theorem \cite{megretski1997system}.
This IQC-based method can handle a broad class of systems, while decomposed stability conditions tend to be conservative, as remarked in the paper. 
System level synthesis (SLS) also enables the modular design of controllers in a discrete-time setting.
The SLS-based method proposed in \cite{wang2018separable} aims at confining disturbance propagation to a local region as making an optimal controller design problem separable.
The plug-and-play method reported in \cite{riverso2016plug} is based on the applicability of model predictive control (MPC) to a class of nonlinear discrete-time systems, where the integration of a distributed MPC with distributed fault detection is considered to realize a localized plug-and-play operation.
We remark that other plug-and-play control methods, such as in \cite{stoustrup2009plug,bendtsen2013plug}, do not necessarily have modularity in design.

We propose a new modular design method of decentralized controllers unlike the abovementioned methods.
Assuming that the network system of interest is initially stable or it has been stabilized, we develop a framework wherein multiple subcontroller designers can individually design and implement their subcontrollers to enhance the ability of local disturbance attenuation, while preserving the stability of the entire network system.
Each subcontroller is designed as a retrofit controller, for which a design method is proposed in \cite{ishizaki2018retrofit}, and the power system applications are reported in \cite{sadamoto2018retrofit,sasahara2019damping}.
The subcontroller is designed as an add-on type subcontroller such that only the model of the subsystem of interest is required during the controller design. 
Robust stability is guaranteed for retrofit control because the resultant feedback system is stable despite variations in the neighboring subsystems, other than the subsystem of interest, assuming that the system is stable prior to implementing the retrofit control.
This stability assumption reflects the fact that ``working" engineering systems in reality, such as power systems, are operated stably as a combination of established techniques.
The proposed modular design method provides a new theoretical basis for sequential system upgrades, such that the stability of the current system is taken over its future generations.

The main contributions in this paper are listed below.
First, considering a single subcontroller designer who manages their own subsystem, we derive a necessary and sufficient condition for the existence of retrofit controllers.
The existence condition is derived in terms of a constrained version of the Youla parameterization \cite{youla1976modern}, 
based on which we also show that the particular structure inside retrofit controllers reported in the previous works \cite{ishizaki2018retrofit,sadamoto2018retrofit,sasahara2019damping} is unique 
if the interaction signal flowing into the subsystem of interest is measurable.
Furthermore we show that simultaneously implementing multiple retrofit controllers that are designed  by different subcontroller designers can contribute to improving the upper bound of the overall control performance, which is defined for the map from all disturbance inputs to all evaluation outputs of the subsystems. 
Finally, its practical significance is demonstrated by an illustrative example of frequency regulation using a standard power system model, known as the IEEE 68-bus power system model.

\footnotetext[1]{
This paper builds on its proceedings versions \cite{sasahara2018characterization,inoue2018parameterization,sasahara2018parameterization}, as collecting the results on the Youla parameterization of retrofit controllers discussed under different settings.
The modular design method of decentralized controllers in this paper is developed based on these preliminary results, for which detailed mathematical proofs are also provided.
Furthermore, a detailed analysis of a power systems application is provided as a numerical example.
Another recent work \cite{ishizaki2019retrofit} presents an advanced technique for determining a better subcontroller in retrofit control.
In particular, assuming that the materials presented in Section~\ref{sec:retro} of this paper are proven facts, a sophisticated technique is developed to employ the partial information regarding environments, assumed to be unknown in this paper.
}

A control system design approach based on passivity, or more generally, dissipativity \cite{willems1972dissipativeIandII,moylan1978stability,qu2014modularized,van2014port}, is relevant to the modular design of network systems.
This approach has an advantage that the input-output behavior of the entire network system can be analyzed by only using those of subsystems, with which compatible supply rates of energy are associated.  
However, analysis based on the supply rates is only valid when an admissible supply rate is found for the joint variable, i.e., the stacked vector, of disturbance and interaction inputs and the joint variable of evaluation and interaction outputs.
Therefore, there is little flexibility in the analysis of general system behavior because a new supply rate for each subsystem must be determined each time if a subcontroller designer reselects the disturbance input and evaluation output ports.
In contrast, our approach, which is based on retrofit control, has higher flexibility in selecting individual input and output ports for subcontroller designers.

The remainder of this paper is organized as follows.
In Section~II, we formulate a modular design problem of decentralized controllers with an example from the control of power systems. 
Thereafter, we provide a formal definition of the retrofit control and perform a detailed analysis of the retrofit control in Section~III, where a constrained version of the Youla parameterization is derived to characterize all retrofit controllers.
Based on the analysis in Section~III, we further analyze the resultant feedback system when multiple retrofit controllers are simultaneously implemented, as described in Section~IV.
Section~V presents the practical significance of our method using an illustrative example of power systems control.
Finally, the conclusions of this paper are presented in Section~VI.

\vspace{2mm}
\noindent \textbf{Notation}~ The notation used in this paper is generally standard.
The identity matrix with an appropriate size is denoted as $I$, the set of stable, proper, real rational transfer matrices as $\mathcal{RH}_{\infty}$, and the $\mathcal{H}_{\infty}$-norm of $G \in \mathcal{RH}_{\infty}$ as $\|G\|_{\infty}$.
All following transfer matrices are assumed to be proper and real rational unless stated otherwise.
A transfer matrix $K$ is said to be a stabilizing controller for $G$ if the feedback system of $G$ and $K$, denoted by $\mathcal{F}(G,K),$ is internally stable in the standard sense \cite{zhou1995robust}.
The vector stacking $v_1,\ldots,v_N$ is denoted as $\sfcol(v_1,\ldots,v_N)$.
The block diagonal matrix whose diagonal entries are $A_1,\ldots,A_N$ is denoted as $\sfdiag(A_1,\ldots,A_N)$ or simply  $\sfdiag(A_i)$.
The real and imaginary parts of a complex number $\bm{z}$ are denoted as $\Re(\bm{z})$ and $\Im(\bm{z})$, respectively.


\section{Preliminaries}\label{sec:prob}

\subsection{Motivating Example from Power Systems Control}\label{secmotex}

\subsubsection{System Description}

As a motivating example, we consider the frequency regulation of the IEEE 68-bus power system model \cite{pal2006robust}, whose network structure is depicted in Fig.~\ref{figieee68}.
This is a standard model employed in bulk power systems; it comprises 16 generators and 35 loads for simulating their oscillatory behavior in response to, e.g., bus faults.
In the following descriptions, the bus connected to a generator is termed as a generator bus, a bus connected to a load is termed as a load bus, and a bus connected to none of them is termed as a non-unit bus.
The label sets of the generator buses, load buses, and non-unit buses are denoted by $\mathcal{I}_{\rm G}$, $\mathcal{I}_{\rm L}$, and $\mathcal{I}_{\rm N}$, respectively.

\begin{subequations}\label{dynpower}
We refer to the dynamics of the $i$th generator as the dynamics of the $i$th generator bus.
The dynamics of the generator buses can be represented as
\begin{equation}\label{gendyn}
\simode{
\dot{x}_i & =f_{i} (x_i, V_i, u_i,U_i)\\
I_i & =g_{i} (x_i, V_i ) \\
\omega_i & =S_i x_i,
}
\quad i\in \mathcal{I}_{\rm G}
\end{equation}
where 
$V_i$ is a two-dimensional real-valued vector composed of the real and imaginary parts of the $i$th bus voltage phasor, 
$I_i$ is a two-dimensional real-valued vector composed of the real and imaginary parts of the $i$th bus outflowing current phasor, 
$x_i$ is a  seven-dimensional generator state composed of two-dimensional swing dynamics with one-dimensional voltage dynamics, one-dimensional excitation system dynamics with an automatic voltage regulator (AVR), and three-dimensional dynamics of a power system stabilizer  (PSS), 
$u_i$ is a scalar reference input to AVR, 
$U_i$ is a scalar reference input for mechanical power regulation, and 
$\omega_i$ is the frequency deviation;  
Additional detail is presented in Appendix~\ref{apx_pmod}.

\begin{figure}[t]
\centering
\includegraphics[width = .95\linewidth]{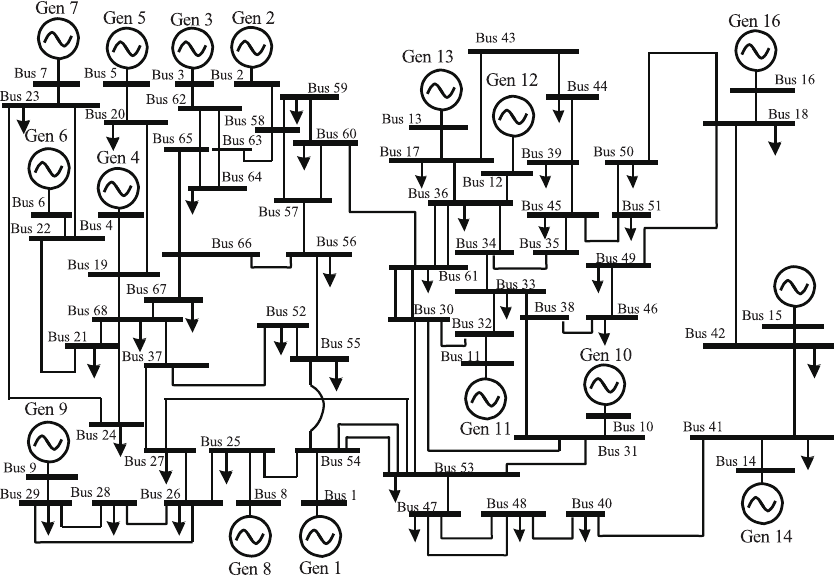}
\caption{IEEE 68-bus test system. Buses, generators, and loads are denoted by bars, circles, and arrows, respectively.}
\label{figieee68}
\end{figure}

The input-output characteristics of the load buses can be represented as a static system.
In this example, we adopt the constant impedance model represented as
\begin{equation}\label{dynload}
I_i = Z^{-1}_i V_i ,\quad i\in \mathcal{I}_{\rm L}
\end{equation}
where 
$V_i$ and $I_i$ are composed of the real and imaginary parts of the $i$th bus voltage and current phasors, and $Z_i$ denotes the load impedance parameter, which is a two-dimensional real-valued nonsingular matrix.
Note that $Z_i$ varies slowly in practice because the loads are variable.
The characteristics of the non-unit buses can be represented as
\begin{equation}\label{dynnon}
I_i =0,\quad i\in \mathcal{I}_{\rm N}.
\end{equation}
The subsystems \eqref{gendyn}--\eqref{dynnon} are interconnected such that 
\begin{equation}\label{powbal}
I_i = \sum_{j=1}^{68} Y_{ij} V_j,\quad i \in \{1,\ldots, 68\}
\end{equation}
where $Y_{ij}$ denotes a two-dimensional real-valued admittance, corresponding to the real and imaginary parts,  associated with the transmission network.
\end{subequations}
The details of these models and the standard values of parameters can be found in \cite{sadamoto2019dynamic}.

\vspace{1mm}
\subsubsection{Automatic Generation Control}

The basic objective of automatic generation control (AGC) is to find a set of suitable mechanical power reference inputs, i.e., $\{U_i\}_{i\in \mathcal{I}_{\rm G}}$, such that all frequency deviations  are sufficiently small.
In practice, the exact data on the load parameters, load characteristics, and transmission network parameters are not available, and these parameters vary slowly on the time scale of the AGC.
Such unknown variations are typically managed via integral-based control \cite[Section~11]{kundur1994power}.
A broadcast-type PI controller that feedbacks the average of the frequency deviations is often used, i.e., 
\begin{equation}\label{gPI}
U_i = - \alpha_i 
\underbrace{
\left(k_{\rm P} \overline{\omega} + k_{\rm I} \int_{0}^{t}\overline{\omega}(\tau)d\tau\right)
}_{\overline{U}}
,\quad 
\overline{\omega}:=\frac{1}{16}\sum_{i =1}^{16} \omega_i
\end{equation}
where $k_{\rm P}$ and $k_{\rm I}$ are nonnegative controller gains, and $\alpha_i$ is a scaling parameter, called a participation factor.
The controller gains are often empirically tuned in an actual operation.
Such empirical gain tuning is possible and justified because the input-output characteristics from $\overline{U}$ to $\overline{\omega}$ in the entire system dynamics represented by \eqref{gendyn}--\eqref{powbal} generally has a passivity-short property for most of the usual operating points,
as will be demonstrated in Section~\ref{sec_nAGC}.
Therefore, in principle, the system stability is attained and explained by the passivity and low-pass properties of the PI controllers.
Therefore, we assume that the AGC maintains the system stability at all operating points of interest in this study.

\vspace{1mm}
\subsubsection{Ground Fault Control}

We focus on the ground fault control (GFC) described here.
First, we explain the ground fault.
The $i$th bus fault, which may occur on any bus, is modeled as a short-time alteration of the system dynamics such that $V_i =0$, which is imposed as an additional physical constraint that causes a non-negligible amount of generator state deviation during the fault.
Typically, the fault duration is approximately 0.1 second or less.
After recovering from the ground fault, the entire system follows the original dynamics before the fault.
The initial state of the system is determined as the deviated state at the moment when the system is recovering from the fault.
We remark that, in general, the ground fault on any bus instantaneously affects the states of all generators; 
this implies that a ground fault can be regarded as a global disturbance that instantly affects all generators.

Although AGC ensures the existence of a stable state space, i.e., a domain of attraction, around each operating point, it is generally not effective for the attenuation of oscillations caused due to ground faults.
This is because the mechanical power reference input $U_i$, which can be used over the time scale of a few seconds, is considerably slower than the time scale of such oscillations.
A possible method of improving the performance of the GFC is to modify or upgrade the originally attached PSSs, each of which shares the same input port as that of the AVR reference input $u_{i}$.
However, such a PSS upgrade is not easy to implement in practice because
\begin{itemize}
\item each PSS designer cannot have exact knowledge about the entire network system, such as load parameters, load characteristics, and transmission network parameters, and 
\item the upgrade of individual PSSs may destroy the entire system stability attained by the AGC, possibly due to unexpected interference among the upgraded PSSs.
\end{itemize}
Therefore, it is crucial to devise a modular design method such that upgrading individual PSSs can contribute to improving the performance of the GFC, while preserving the system stability achieved by the AGC.
In fact, a simple decentralized control method for PSS upgrade easily destabilizes the network system, as will be demonstrated in Section~\ref{secnum}.

Regarding the PSS upgrade, we assume that each PSS designer assigned to a corresponding generator can only have their own generator model \eqref{gendyn}, or its linearized version 
\begin{equation}\label{lingendyn}
\simode{
\dot{x}_i& =A_i^{\star} x_i + L_i^{{\star}} V_i +B_i^{{\star}} u_i +R_{i}^{{\star}} U_i \\
I_i & = \mathit{\Gamma}_i^{{\star}} x_i + D_i^{{\star}} V_i\\
\omega_i & = S_i x_i,
}
\quad i\in \mathcal{I}_{\rm G},
\end{equation}
where the system matrices with ``$\star$" in \eqref{lingendyn} are dependent on an operating point of interest, and all variables are redefined as the deviations from the operating point.
The operating point of each generator, denoted by $\{x_i^{\star}, V_i^{\star}, U_i^{\star}\}$, is an implicit function of the load impedance set $\{Z_i\}_{i \in \mathcal{I}_{\rm L}}$, and it satisfies
\[
f_i (x_i^{\star},V_i^{\star},0,U_i^{\star})=0,\quad S_i x_i^{\star}=0
\]
for all the generators.
The linearized model \eqref{lingendyn} can be derived from \eqref{gendyn} because each PSS designer can identify the operating point at which the system is currently driven, as a result of monitoring the actual behavior of local physical variables.
This is enabled by the fact that the dynamics of the AGC is considerably slower than that of GFC, over the time scale of which the assumption of quasi-steady states is generally valid.

\subsection{Problem Formulation}\label{subsecprob}
\subsubsection{System Description}

\begin{figure}[t]
\centering
\includegraphics[width = .75\linewidth]{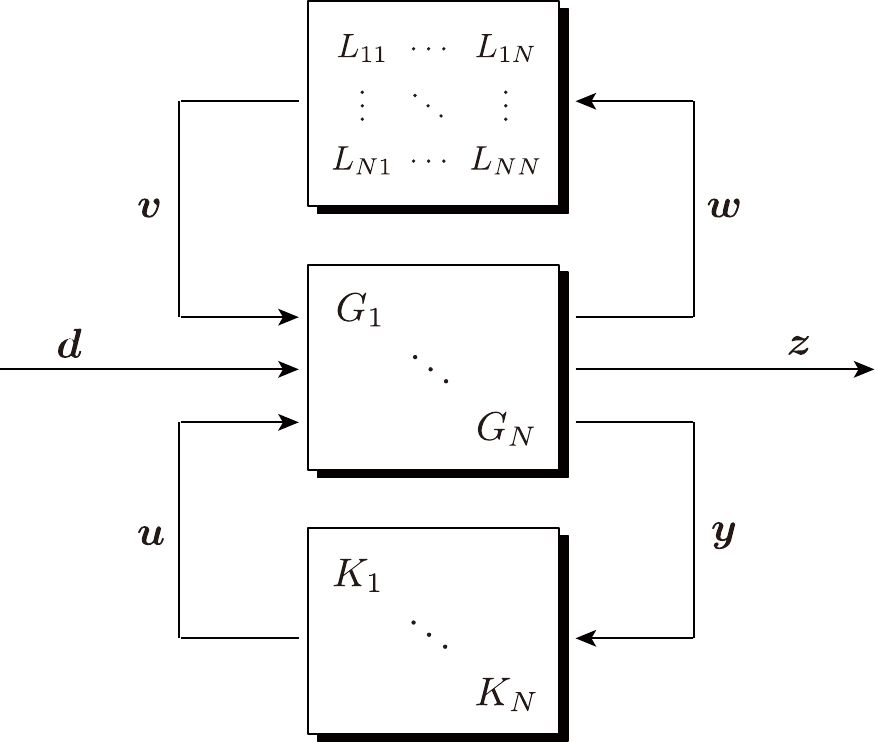}
\caption{The network system comprising $N$ subsystems. 
The middle block represents the subsystems, top block represents the interaction among the subsystems, and bottom block represents subcontrollers.
}
\label{fig:ent_N}
\end{figure}

Considering the upgrade of PSSs described in Section~\ref{secmotex}, we formulate a modular design problem of decentralized controllers, assuming linearization at an operating point of interest.
Consider the network system depicted in Fig.~\ref{fig:ent_N}.
Denoting the components of each signal in the block diagram, e.g., as
\[
\bm{w}=\sfcol(w_1,\ldots,w_N),\quad
\bm{v}=\sfcol(v_1,\ldots,v_N),
\]
we describe the dynamics of the $i$th subsystem as
\begin{equation}\label{subsysG}
\mat{
 w_i\\
 z_i\\
 y_i
}
 = 
\underbrace{
\mat{
  G_{w_i v_i} & G_{w_i d_i} & G_{w_i u_i}\\
  G_{z_i v_i} & G_{z_i d_i} & G_{z_i u_i}\\
  G_{y_i v_i} & G_{y_i d_i} & G_{y_i u_i}
}
 }_{G_i}
\mat{
 v_i\\
 d_i\\
 u_i
}
\end{equation}
where $v_i$ and $w_i$ are the interaction input and output, $d_i$ and $z_i$ are the disturbance input and evaluation output, and $u_i$ and $y_i$ are the control input and measurement output, respectively.
The interaction among the subsystems is expressed as
\begin{equation}\label{intL}
\mat{
  v_1\\
  \vdots\\
  v_N
}
 =
 \underbrace{
\mat{
  L_{11} & \cdots & L_{1N}\\
  \vdots & \ddots & \vdots\\
  L_{N1} & \cdots & L_{NN}
}
}_{\bm{L}}
\mat{
  w_1\\
  \vdots\\
  w_N
},
\end{equation}
which can also be a dynamical system.
Similarly, the decentralized controller is expressed as
\begin{equation}\label{decK}
\mat{
  u_1\\
  \vdots\\
  u_N
}
 =
\underbrace{
\mat{
  K_{1} & & \\
  & \ddots & \\
 &  & K_{N}
}
}_{\bm{K}}
\mat{
  y_1\\
  \vdots\\
  y_N
}.
\end{equation}
For simplicity of discussion, throughout this paper, we assume that all feedback systems are well-posed.
The main goal of this study is to devise a modular design method to determine a decentralized controller $\bm{K}$ such that the effect of the disturbance input $\bm{d}$ on the evaluation output $\bm{z}$ is reduced.

The object to be controlled is the network system composed of $G_1,\ldots, G_N$ interacted by $\bm{L}$.
The diagonally stacked versions of the transfer matrices in \eqref{subsysG} are denoted as, e.g.,
\begin{equation}\label{diagG}
\bm{G}_{\bm{yu}}:=\sfdiag(G_{y_i u_i}),\quad
\bm{G}_{\bm{wv}}:=\sfdiag(G_{w_iv_i}).
\end{equation}
Subsequently, we refer to the feedback system composed of the blocks of $\bm{L}$ and $\bm{G}$ in Fig.~\ref{fig:ent_N}, i.e.,
\begin{equation}\label{platsys}
\bm{G}_{\rm pre}:= \mathcal{F}(\bm{L},\bm{G})
\end{equation}
as a \textit{preexisting system}.
Using this notation, we assume the following throughout this paper.

\begin{assumption}\label{assentsys}
For the network system in Fig.~\ref{fig:ent_N}, assume that
\begin{itemize}
\item[\textbf{(a)}] the preexisting system $\bm{G}_{\rm pre}$ is internally stable, and
\item[\textbf{(b)}]  each subcontroller $K_i$ is  designed by a corresponding subcontroller designer who only possesses the model information about his or her subsystem $G_i$.
\end{itemize}
\end{assumption}

Regarding the GFC mentioned in Section~\ref{secmotex}, Assumption~\ref{assentsys}\textbf{(a)} reflects the fact that the stability of an equilibrium of interest is attained by the AGC, and Assumption~\ref{assentsys}\textbf{(b)} implies the modularity in the upgrade of each PSS.
We can regard $G_i$ as the $i$th generator dynamics, which is seven-dimensional and linearized, and $K_i$ as an ``upgrade module" designed for individual PSS to improve the performance of GFC.
Note that the dynamics of $\bm{L}$ is supposed to encapsulate the broadcast-type PI controller in \eqref{gPI} in addition to the algebraic load characteristics in \eqref{dynload}, non-unit buses in \eqref{dynnon}, and  interconnection in \eqref{powbal}, where unknown system parameters are involved.
The interaction output signal $w_i$ is identified as $\sfcol(I_i,\omega_i)$ and the interaction input signal $v_i$ as $\sfcol(V_i,U_i)$.
The measurement output $y_i$ can be selected as a part or all of the state variables of the $i$th generator,   evaluation output $z_i$ is to be selected as the frequency deviation $\omega_i$, and disturbance input $d_i$ abstracts the effect of ground faults.

\vspace{1mm}
\subsubsection{Modular Subcontroller Design Problem}

\begin{figure}[t]
\centering
\includegraphics[width = 0.50\linewidth]{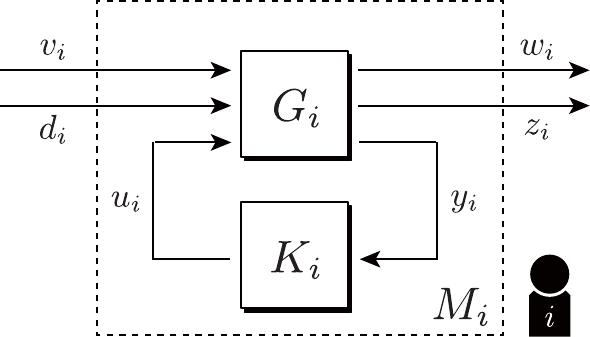}
\caption{
Modular design of a subcontroller by the $i$th designer.}
\label{figmodsys}
\end{figure}

Subsequently, we formulate a design problem of the decentralized controller based on Assumption~\ref{assentsys}.
For the ``isolated" feedback system shown in Fig.~\ref{figmodsys}, we introduce the map
\begin{equation}\label{modcon}
M_i:(v_i,d_i)\mapsto (w_i,z_i).
\end{equation}
A naive subcontroller design problem may be written as a problem of finding a subcontroller $K_i$ that stabilizes the isolated feedback system $M_i$.
However, such individual subcontroller designs may destroy the stability of the entire network system due to unexpected interference among the subcontrollers.
A possible method to avoid this interference is to formulate a ``constrained" subcontroller design problem in the form of
\begin{equation}\label{conopt}
J_i \bigl[M_i (K_i)\bigr]\leq J_i^{\star},\quad K_i  \in \mathcal{M}_i
\end{equation}
where $J_i$ denotes an objective function, $J^{\star}_i$ denotes its admissible bound, and $\mathcal{M}_i$ denotes a set of subcontrollers complying with a desirable stability requirement.
Using this notation, we address the following problem.

\begin{problem}\label{probmain}
Let Assumption~\ref{assentsys} hold.
Then, find a family of the subcontroller sets $\mathcal{M}_1,\ldots,\mathcal{M}_N$ such that the entire network system in Fig.~\ref{fig:ent_N} is internally stable for any choice of a tuple
\begin{equation}\label{famm}
(K_1,\ldots, K_N)  \in \mathcal{M}_1 \times \cdots \times \mathcal{M}_N.
\end{equation}
Furthermore, determine a set of the individual objective functions $J_1,\ldots,J_N$ such that the $\mathcal{H}_{\infty}$-norm of the entire  system map $\bm{T}_{\bm{zd}} :\bm{d}\mapsto \bm{z}$ is bounded as
\begin{equation}\label{bTzdbnd}
\|\bm{T}_{\bm{zd}} \|_{\infty}\leq \gamma(J_1^{{\star}},\ldots,J_N^{{\star}}),
\end{equation}
where $\gamma:\mathbb{R}^{N}\rightarrow \mathbb{R}$ is a monotonically increasing function for each argument.
\end{problem}

We refer to this problem as a \textit{modular design problem of decentralized controllers}, in which a family of $\mathcal{M}_1,\ldots,\mathcal{M}_N$ can be regarded as a class of  decentralized controllers that can preserve the system stability assumed as Assumption~\ref{assentsys}\textbf{(a)}.
Each subcontroller is assumed to be individually designed considering Assumption~\ref{assentsys}\textbf{(b)}.
The existence of a monotonically increasing function $\gamma$ ensures that individual designs of multiple subcontrollers can contribute to improving at least the upper bound of the overall control performance.
More specifically, the smaller the values of $J_1^{\star},\ldots,J_N^{\star}$, the smaller is the upper bound.
In a practical situation, each subcontroller designer is expected to minimize their performance bound $J_i^{\star}$.
A particular form of $\gamma$ is presented in Section~\ref{sec:2stage}.
It should be noted that most of the discussions in this paper are not limited to the evaluation based on the $\mathcal{H}_{\infty}$-norm. 
Other measures such as the $\mathcal{H}_{2}$-norm can also be used in a similar manner.

\begin{figure*}[t]
\centering
\includegraphics[width = .75\linewidth]{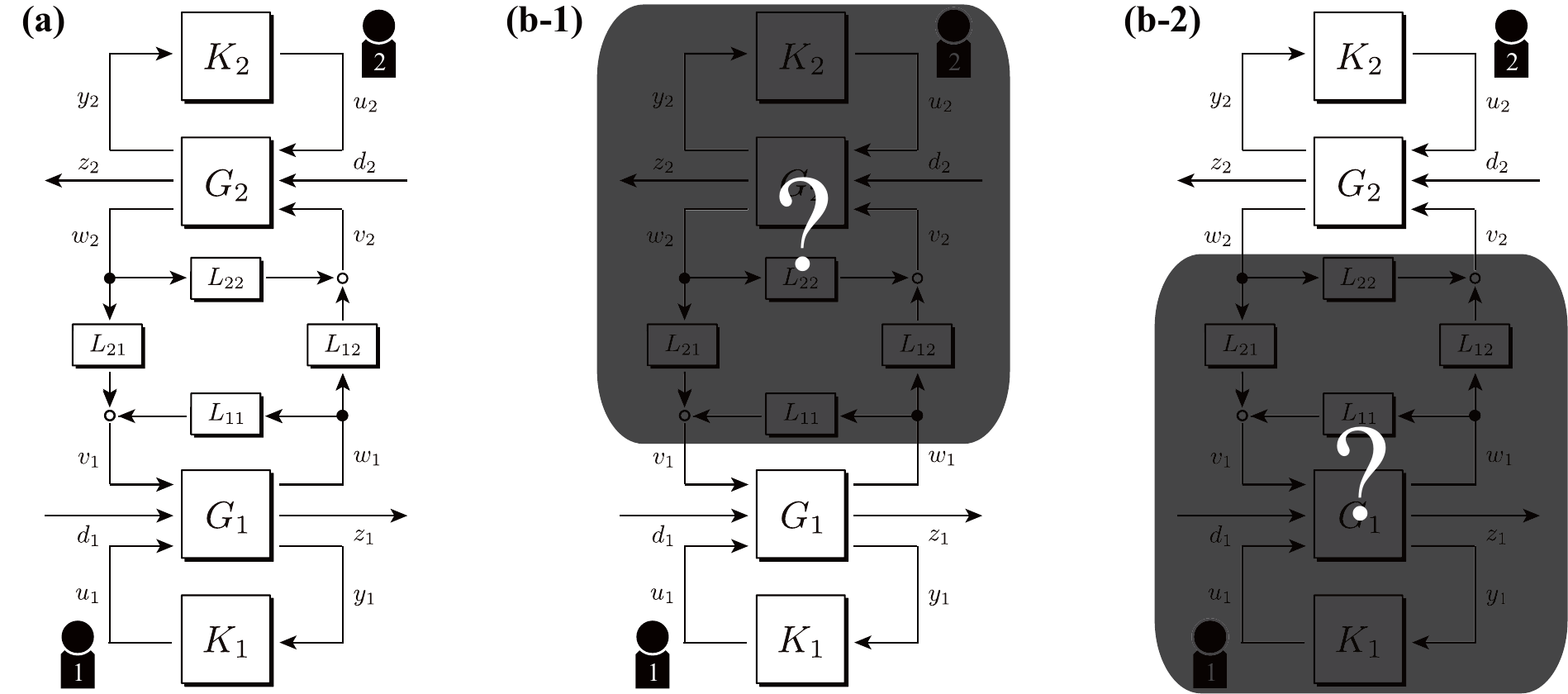}
\caption{Example of modular design of decentralized controllers with two subsystems.
\textbf{(a)}: Overview. 
\textbf{(b-1)}: View from the perspective of the first subcontroller designer.
\textbf{(b-2)}: View from the perspective of the second subcontroller designer.
}
\label{fig:ex}
\end{figure*}


\begin{figure}[t]
\centering
\includegraphics[width = .55\linewidth]{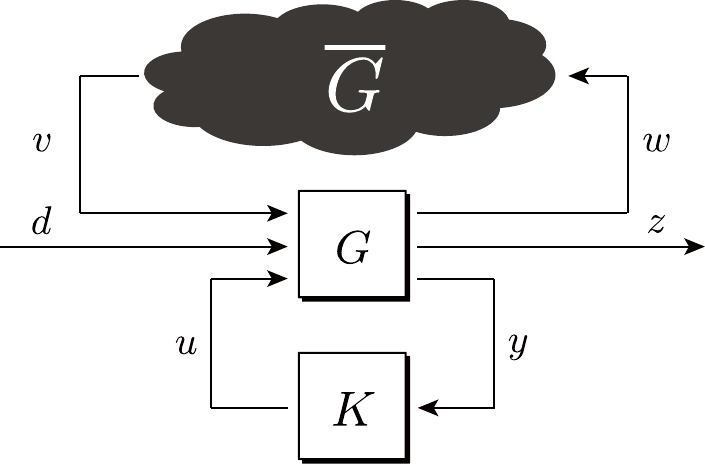}
\caption{Retrofit control from the perspective of each subcontroller designer. 
The block of $K$ represents the retrofit controller to be designed, and the blocks of $G$ and $\overline{G}$ represent the subsystem of interest and its unknown environment, respectively.}
\label{fig:pre_sys}
\end{figure}

\section{Analysis from Perspective of Single Designer}\label{sec:retro}

\subsection{Definition of Retrofit Control}

In this section, we first analyze the modular design problem of decentralized controllers from the perspective of ``one'' subcontroller designer, for whom the information regarding the interaction, other subsystems and subcontrollers is supposed to be concealed.
To explain this further, we consider an example using the two subsystems illustrated in Fig.~\ref{fig:ex}, where Fig.~\ref{fig:ex}\textbf{(a)} presents the overview, and Figs.~\ref{fig:ex}\textbf{(b-1)} and \textbf{(b-2)} present local views from the perspectives of the first and second subcontroller designers, respectively.
As depicted in the figure, each designer aims at designing their subcontroller while considering the remainder of the network system as their ``unknown" environment.
Such a control problem is termed as a retrofit control problem, and several controller design methods for this problem have been reported in \cite{ishizaki2018retrofit,sadamoto2018retrofit,sasahara2019damping}.

The abstraction of retrofit control is depicted in Fig.~\ref{fig:pre_sys}, where the subsystem and subcontroller of one designer are denoted by $G$ and $K$, respectively, and their environment is denoted by $\overline{G}$.
Throughout this section, we regard Fig.~\ref{fig:pre_sys} as the feedback system from the standpoint of the $i$th designer, while neglecting the subscript ``$i$" for simplicity of notation.
For the subsequent discussion, we use symbols denoting the submatrices of $G$, for example, as
\[
\spliteq{
G_{(z,y)(d,u)}&:=
\mat{
G_{zd}&G_{zu}\\
G_{yd}&G_{yu}
},
\quad
G_{(z,y)v}:=
\mat{
G_{zv}\\
G_{yv}
},
\\
G_{w(d,u)}&:=
\mat{
G_{wd}&G_{wu}\\
}.
}
\]
Furthermore, we define the feedback system
\begin{equation}\label{presys}
G_{\rm pre}:=\mathcal{F}(\overline{G},G),
\end{equation}
which is a rewrite of $\bm{G}_{\rm pre}$ in \eqref{platsys} from the perspective of the subcontroller designer of interest.
Using this notation, we define the following notion of retrofit controllers.

\begin{definition}\label{def:retro}
Define the set of all admissible environments as
\[
\overline{\mathcal{G}} := \{ \overline{G}: G_{\rm pre}\ {\rm is\ internally\ stable}\}.
\]
An output feedback controller 
\[
u=Ky
\] 
is said to be a \textit{retrofit controller} if the entire feedback system in Fig.~\ref{fig:pre_sys} is internally stable for any environment $\overline{G} \in \mathcal{\overline{G}}$.
\end{definition}

The retrofit controller is defined as an add-on type subcontroller that can ensure the stability of the resultant feedback system for any possible variation of environments such that the preexisting system is stable.
The stability of the preexisting system is based on Assumption~\ref{assentsys}\textbf{(a)}.
Within the set of all such retrofit controllers, each subcontroller designer aims at selecting a desirable subcontroller that can improve the resultant control performance for their local disturbance attenuation.
Although the environment in this formulation may be regarded as model uncertainty in robust control, it is typically assumed to be ``norm-bounded" in a standard robust control setting. 
It is evident that we do not impose any explicit norm bound on the environment.
In this sense, the retrofit control problem is different from standard robust control problems.

\begin{figure}[t]
\centering
\includegraphics[width = .60\linewidth]{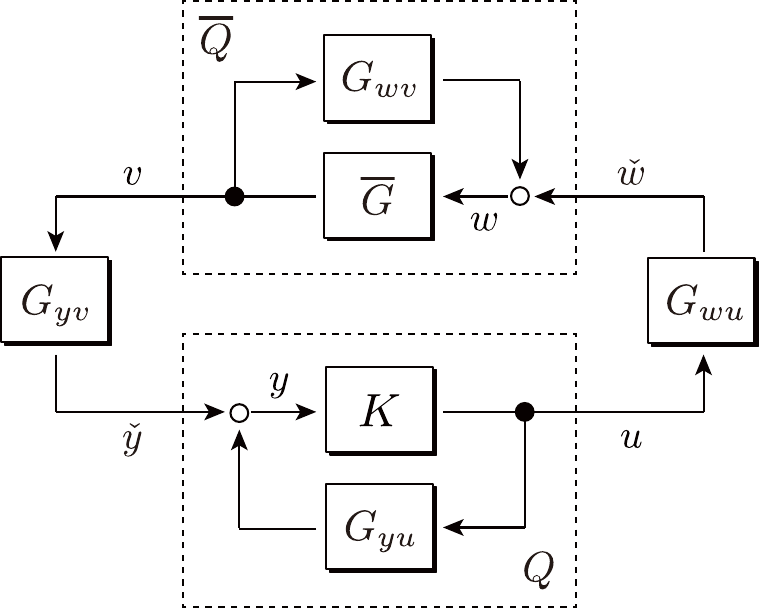}
\caption{Block diagram obtained by the Youla parameterization.}
\label{fig:loop_sys}
\end{figure}

\subsection{Parameterization of All Retrofit Controllers}

In this subsection, as one of the main contributions of this paper, we describe the parameterization of ``all" retrofit controllers, which involves both classes of retrofit controllers reported in~\cite{ishizaki2018retrofit} as special cases.
For simplicity of presentation, we first make the following assumption.

\begin{assumption}\label{assum:loc_stable}
The subsystem $G$  is stable, i.e., $G\in\mathcal{RH}_{\infty}$.
\end{assumption}

Assumption~\ref{assum:loc_stable} is not essential to prove the results below, but it is made here to maintain their readability as the Youla parameterization \cite{youla1976modern} is simplified for controller parameterization.
Generalizations to unstable subsystems are presented in Appendix~\ref{app:general}.
As shown in the following theorem, all retrofit controllers can be parameterized as a constrained version of the Youla parameterization.

\begin{theorem}\label{thm:main}
Let Assumption~\ref{assum:loc_stable} hold.
Consider that the Youla parameterization of $K$ is given as
\begin{equation}\label{eq:youK}
 K = (I+QG_{yu})^{-1}Q,\quad Q\in \mathcal{RH}_{\infty}
\end{equation}
where $Q$ denotes its Youla parameter.
Then, $K$ is a retrofit controller if and only if
\begin{equation}\label{eq:gencond_sta}
 G_{wu}QG_{yv}=0.
\end{equation}
\end{theorem}

A complete proof can be found in Appendix~\ref{sec:pfthm}.
The essence of the proof, which provides an interpretation of the constraint \eqref{eq:gencond_sta}, is explained as follows.
Let us drop all the terms relevant to $d$ and $z$ in Fig.~\ref{fig:pre_sys}, because they are not essential for proving the internal stability of the resultant feedback system.
On denoting the Youla parameterization of $\overline{G}$ as
\begin{equation}\label{eq:youGE}
\overline{G} = (I+\overline{Q}G_{wv})^{-1}\overline{Q},\quad \overline{Q}\in \mathcal{RH}_{\infty},
\end{equation}
we obtain the closed-loop system depicted in Fig.~\ref{fig:loop_sys}, which is composed of the feedback of $\overline{Q}$ and $G_{wu}QG_{yv}$.
Note that $\overline{Q}$ can be considered as an arbitrary element in $\mathcal{RH}_{\infty}$ because $\overline{G}$ is assumed to be arbitrary in $\overline{\mathcal{G}}$.
Therefore, \eqref{eq:gencond_sta} is shown to be necessary and sufficient for the internal stability.
We can also see from Fig.~\ref{fig:loop_sys} that $Q: \check{y} \mapsto u$ and $\overline{Q} : \check{w}\mapsto v$ where $\check{y}$ and $\check{w}$ are different from $y$ and $w$, respectively.

It is noteworthy that all retrofit controllers are characterized as a set of subcontrollers that maintain a constant ``interaction transfer matrix," which is as follows.
Let $M_{vw}$ denote the map from $v$ to $w$ in Fig.~\ref{figmodsys}, i.e.,
\[
M_{wv}:= G_{wv} +  G_{wu} \underbrace{K(I-G_{yu}K)^{-1}}_{Q} G_{yv}.
\]
It is clear that \eqref{eq:gencond_sta} is necessary and sufficient for $M_{wv}=G_{wv}$, implying that the retrofit controller can be characterized as a controller such that it does not change the dynamics from the interaction input $v$ to the interaction output $w$.
The principle behind the retrofit control can also be explained in terms of decoupling control, or disturbance rejection (interconnection signal rejection) control \cite{falb1967decoupling,wang2013distributed} applied to the ``transformed system representation" derived by the Youla parameterization, instead of the original system representation.

Based on the constrained Youla parameterization in Theorem~\ref{thm:main}, we clarify a structure inside the entire feedback system map.
Let $T_{zd}:d \mapsto z$
denote the entire map compatible with  Fig.~\ref{fig:pre_sys}.
Then, for any $K$ that is not necessarily a retrofit controller, we have
\[
\spliteq{
 T_{zd} &  = \mathcal{F}_{\rm u} \left( \mathcal{F}_{\rm l} \left(G,K \right),\overline{G} \right) \\
& = \mathcal{F}_{\rm u}\left(
\mathcal{F}_{\rm l}
\left(
\mat{
  0 & G_{wd} & G_{wu}\\
  G_{zv} & G_{zd} & G_{zu}\\
  G_{yv} & G_{yd} & 0
}, 
Q
\right), 
\overline{Q}
\right) \\
  &  = M_{zd}(Q)\!+\!M_{zv}(Q)
\underbrace{
  (I-\overline{Q}G_{wu}QG_{yv})^{-1}
}_{\star}
\overline{Q}M_{wd}(Q)
}
\]
where $ \mathcal{F}_{\rm u}$ and $ \mathcal{F}_{\rm l}$ denote the lower and upper linear fractional transformations \cite{zhou1995robust}, respectively.
Additionally,
\begin{equation}\label{trmM}
 \spliteq{
 M_{zd}(Q) &  := G_{zd}+G_{zu}QG_{yd}, \\
 M_{zv}(Q) &  := G_{zv}+G_{zu}QG_{yv}, \\
 M_{wd}(Q) &  := G_{wd}+G_{wu}QG_{yd}.
 }
\end{equation}
These transfer matrices correspond to the lower three blocks of $M$ in \eqref{modcon}, i.e.,
\[
M(Q)=  
\mat{
  M_{wv}(Q) & M_{wd}(Q)\\
  M_{zv}(Q) & M_{zd}(Q)
}.
\]
Recall that $ M_{wv}=G_{wv}$ for all retrofit controllers.
Furthermore, the feedback term represented by ``$\star$" is reduced to the identity matrix.
This proves the following fact.

\begin{theorem}\label{thm:cl}
Let Assumption~\ref{assum:loc_stable} hold.
For a retrofit controller $K$ in Theorem~\ref{thm:main}, it follows that
\begin{equation}\label{eq:clmap}
T_{zd}(Q) = M_{zd}(Q)+M_{zv}(Q)\overline{Q}M_{wd}(Q).
\end{equation}
\end{theorem}

Theorem~\ref{thm:cl} shows that $T_{zd}$ is affine with respect to $\overline{Q}$ in retrofit control.
Because $\overline{Q}$ is assumed to be unavailable, for performance regulation, we can only determine a Youla parameter $Q$ subject to the constraint \eqref{eq:gencond_sta} such that the magnitudes of $M_{zd}$, $M_{zv}$, and $M_{wd}$ are jointly reduced.
However, the constraint on $Q$ cannot directly be handled by a standard controller design technique.
Based on \cite[Fact~6.4.43]{bernstein2009matrix}, such a $Q$ may be written as
\[
 Q = Q_0-G_{wu}^{\dagger}G_{wu}Q_0G_{yv}G_{yv}^{\dagger},\quad Q_0 \in \mathcal{R}\mathcal{H}_{\infty}
\]
where $G_{wu}^{\dagger}$ and $G_{yv}^{\dagger}$ denote the right-inverse and left-inverse of $G_{wu}$ and $G_{yv}$, respectively.
However, $G_{wu}^{\dagger}$ and $G_{yv}^{\dagger}$ here are not always found, especially over the ring of $\mathcal{R}\mathcal{H}_{\infty}$.
In this sense, finding a retrofit controller in this general setting is not tractable and straightforward.

\subsection{Tractable Class of Retrofit Controllers}\label{sectract}

In this subsection, we propose a retrofit controller that can be easily designed using a standard controller design technique.
To this end, we introduce the following class of retrofit controllers based on the characterization in Theorem~\ref{thm:main}.

\begin{definition}\label{def:out}
Let Assumption~\ref{assum:loc_stable} hold.
Then, $K$ is said to be an \emph{output-rectifying retrofit controller} if
\begin{equation}\label{eq:condyv}
 QG_{yv} = 0
\end{equation}
where $Q$ denotes the Youla parameter of $K$ in Theorem~\ref{thm:main}.
\end{definition}

It is clear that \eqref{eq:condyv} is a sufficient condition for the constraint \eqref{eq:gencond_sta}.
We remark that confining our attention to \eqref{eq:condyv} does not lose the generality for the power system example in Section~\ref{secmotex}.
This can be seen below.
A necessary condition for the existence of a nonzero $Q$ such that \eqref{eq:gencond_sta} holds is that at least either the dimension of $u$ is strictly larger than that of $w$ or the dimension of $y$ is strictly larger than that of $v$, except for the special cases where $G_{wu}$ or $G_{yv}$ is rank deficient.
For the example in Section~\ref{secmotex}, the dimension of $w$, composed of the real and imaginary parts of the bus current phasor and frequency deviation, is larger than that of the scalar input $u$, denoting a reference signal to the AVR.
This implies that the right kernel of $G_{wu}$ other than zero is null.
As seen here, confining our attention to \eqref{eq:condyv} is reasonable, or possibly necessary, in the case where the number of available input ports is limited due to factors such as physical or economic limitations of actuation.

In the following discussion, we describe the reason for terming as ``output-rectifying"  by deriving an explicit representation of all such retrofit controllers, which clarifies a particular structure inside them. 
We assume the following situation throughout this subsection.

\begin{assumption}\label{assum:v}
The interaction signal $v$ is measurable in addition to the measurement output $y$.
\end{assumption}

From a symbolic perspective, Assumption~\ref{assum:v} corresponds to the situation where every symbol $y$ in the above discussion is to be replaced with the augmented measurement output $(y,v)$.
Based on this assumption, the transfer matrices in \eqref{subsysG} relevant to $y$ are also augmented.
For example, $G_{yv}$ and $G_{yu}$ are to be replaced with
\begin{equation}\label{eq:notation}
 G_{(y,v)v} := 
\mat{
 G_{yv}\\
 I
},
 \quad
 G_{(y,v)u} := 
\mat{
 G_{yu}\\
 0
}.
\end{equation}
Furthermore, the controller $K$  is also augmented as
\begin{equation}\label{eq:outctrl}
 u = K
 \mat{
 y\\
 v
}.
\end{equation}
Subsequently, the Youla parameterization of this $K$ is given by
\begin{equation}\label{eq:cond0}
K=(I+QG_{(y,v)u})^{-1}Q,\quad Q\in \mathcal{RH}_{\infty},
\end{equation}
and its constraint corresponding to \eqref{eq:condyv} is written as
\begin{equation}\label{eq:cond1}
 QG_{(y,v)v} = 0.
\end{equation}
The aim of Assumption~\ref{assum:v} is to enable the following factorization of $Q$ such that \eqref{eq:cond1} holds over the ring of $\mathcal{RH}_{\infty}$.

\begin{lemma}\label{lemfacQ}
Let Assumptions~\ref{assum:loc_stable} and~\ref{assum:v} hold.
Then, the Youla parameter $Q\in \mathcal{RH}_{\infty}$  satisfies \eqref{eq:cond1} if and only if there exists $\hat{Q}\in \mathcal{RH}_{\infty}$ such that $Q=\hat{Q}R$ where
\begin{equation}\label{outrec}
 R:=
\mat{
 I & -G_{yv}
}.
\end{equation}
\end{lemma}

\begin{IEEEproof}
The ``if" part can be easily proven as $R \in \mathcal{RH}_{\infty}$ and $RG_{(y,v)v}=0$.
The ``only if" part is proven as follows.
We apply the calculus over the ring of  $\mathcal{RH}_{\infty}$.
Consider
\[
U:=
\mat{
 I & -G_{yv}\\
 0&I
}.
\]
This $U$ is unimodular, i.e., it is invertible in $\mathcal{RH}_{\infty}$.
Thus, for any $Q \in \mathcal{RH}_{\infty}$, there exists $\tilde{Q}\in \mathcal{RH}_{\infty}$ such that $Q=\tilde{Q}U$.
Substituting this into \eqref{eq:cond1}, we have
\[
\underbrace{
\mat{
\tilde{Q}_{1} & \tilde{Q}_{2}
}
 }_{\tilde{Q}}
\mat{
 I & -G_{yv}\\
 0&I
}
\mat{
 G_{yv}\\
 I
}
=0,
\]
which is equivalent to $\tilde{Q}_2 =0 $.
Note that the upper half of $U$ is equal to $R$.
Hence, for any $Q \in \mathcal{RH}_{\infty}$, there exists some $\tilde{Q}_1 \in \mathcal{RH}_{\infty}$ such that $Q=\tilde{Q}_1 R$.
\end{IEEEproof}

\begin{figure}[t]
\centering
\includegraphics[width = .60\linewidth]{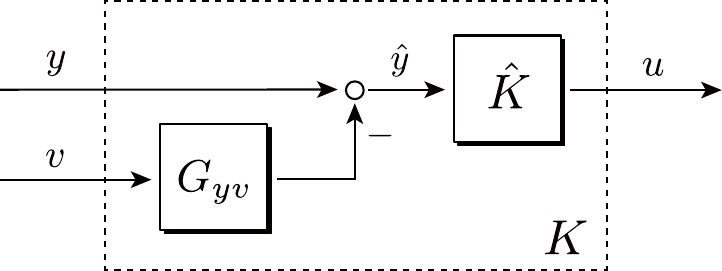}
\caption{Structure inside all output-rectifying retrofit controllers when interaction signal is measurable.}
\label{fig:retro_str}
\end{figure}

Lemma~\ref{lemfacQ} derives a compact expression of  ``all" $Q \in \mathcal{RH}_{\infty}$ such that  \eqref{eq:cond1} holds based on Assumption~\ref{assum:v}.
We notice that $R$ in \eqref{outrec} corresponds to a basis of the left kernel of $G_{(y,v)v}$ in $\mathcal{RH}_{\infty}$.
Thus,  $\hat{Q}$ can be regarded as a new component in the basis of $R$.
Using the factorization in Lemma~\ref{lemfacQ}, we can rewrite \eqref{eq:cond0} as
\begin{equation}\label{youKh}
K=
\underbrace{
(I+\hat{Q}G_{yu})^{-1}\hat{Q}
}_{\hat{K}}R,
\quad\hat{Q}\in \mathcal{RH}_{\infty},
\end{equation}
where we have used the fact that
\begin{equation}\label{GRG}
RG_{(y,v)u}=G_{yu}.
\end{equation}
From \eqref{youKh}, we find that $\hat{K}$ is a stabilizing controller for $G_{yu}$, and $\hat{Q}$ is its Youla parameter.
We refer to 
$R:(y,v)\mapsto \hat{y}$
as an \textit{output rectifier}; this term is used because the measurement output $(y,v)$ is rectified such that
\[
\hat{y}=y-G_{yv}v.
\]
This output rectifier, corresponding to the basis $R$ in \eqref{outrec}, can be regarded as a dynamical simulator to reduce the interference of $v$ with the output signal $y$,  which forwards the rectified output $\hat{y}$ to the stabilizing controller $\hat{K}$.
This discussion leads to the following ``explicit" parameterization of the output-rectifying retrofit control with the interaction signal measurement.

\begin{proposition}\label{thm:out_retro}
Let Assumptions~\ref{assum:loc_stable} and~\ref{assum:v} hold.
Then, $K$ is an output-rectifying retrofit controller if and only if 
\begin{equation}\label{eq:out_retro}
 K = \hat{K}R
\end{equation}
where $\hat{K}$ is a stabilizing controller for $G_{yu}$, i.e., all such retrofit controllers have the structure shown in Fig.~\ref{fig:retro_str}.
\end{proposition}

Subsequently, we analyze the entire feedback system map from the disturbance input to the evaluation output, when we apply the output-rectifying retrofit control in Proposition~\ref{thm:out_retro}.
This analysis is performed based on Theorem~\ref{thm:cl}.
Under Assumption~\ref{assum:v}, the transfer matrices in \eqref{trmM} are augmented as
\[
\spliteq{
 M_{zd}(Q) & = G_{zd}+G_{zu}QG_{(y,v)d},\\
 M_{zv}(Q) & = G_{zv}+G_{zu}QG_{(y,v)v},\\
 M_{wd}(Q) & = G_{wd}+G_{wu}QG_{(y,v)d}.
}
\]
The factorization of $Q$ in Lemma~\ref{lemfacQ} enables the reduction of
$RG_{(y,v)v}=0$ and $RG_{(y,v)d}=G_{yd}$, implying that $M_{zv}$ is equal to $G_{zv}$, and $M_{zd}$ and $M_{wd}$ are, respectively, equal to
\begin{equation}\label{hatMs}
\spliteq{
\hat{M}_{zd}(\hat{K})&:= G_{zd}+G_{zu}\hat{K}(I-G_{yu}\hat{K})^{-1}G_{yd} \\
\hat{M}_{wd}(\hat{K})&:= G_{wd}+G_{wu}\hat{K}(I-G_{yu}\hat{K})^{-1}G_{yd},
}
\end{equation}
where we have  plugged-in the Youla parameter
\[
\hat{Q}= \hat{K}(I-G_{yu}\hat{K})^{-1}.
\]
Using this notation, we have the following result.

\begin{figure}[t]
\centering
\includegraphics[width = .90\linewidth]{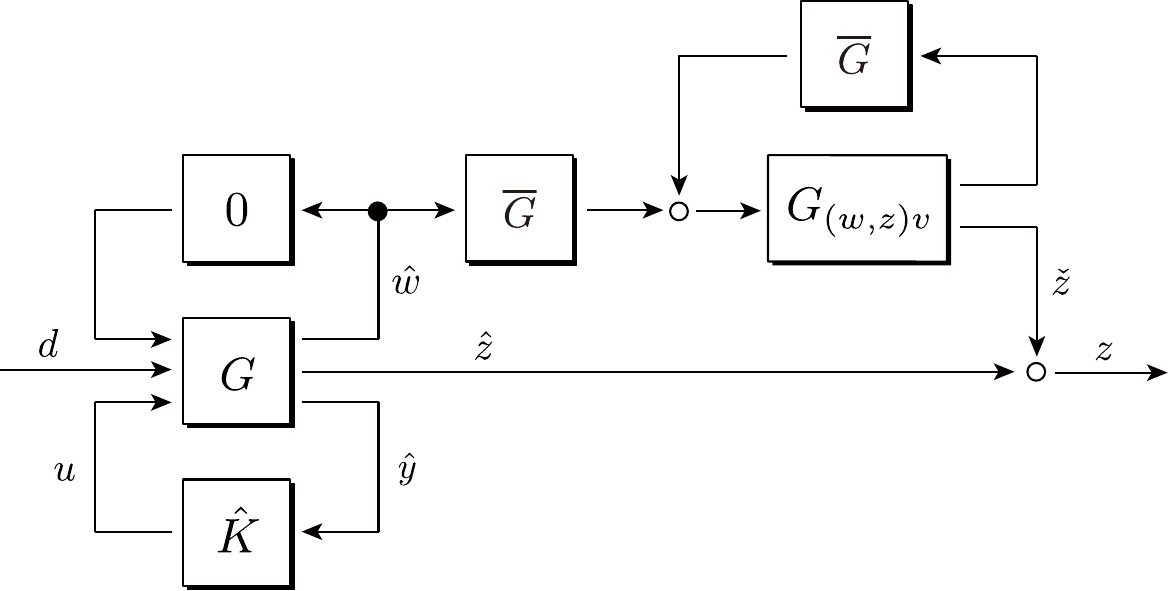}
\caption{Block diagram of output-rectifying retrofit control.}
\label{fig:T_out}
\end{figure}

\begin{proposition}\label{co:cl_out}
Let Assumptions~\ref{assum:loc_stable} and~\ref{assum:v} hold.
For an output-rectifying retrofit controller $K$ in Proposition~\ref{thm:out_retro}, it follows that
\begin{equation}\label{retTzd}
T_{zd}(\hat{K})=\hat{M}_{zd}(\hat{K})+G_{zv}(I-\overline{G}G_{wv})^{-1}\overline{G}\hat{M}_{wd}(\hat{K}),
\end{equation}
and its block diagram is depicted in Fig.~\ref{fig:T_out}.
\end{proposition}

Proposition~\ref{co:cl_out} shows that the block diagram in Fig.~\ref{fig:pre_sys} can be equivalently transformed into that shown in Fig.~\ref{fig:T_out}, when $K$ is an output-rectifying retrofit controller with the interaction signal measurement.
Note that  Fig.~\ref{fig:T_out} has a cascade structure, where the upstream feedback system is composed of the blocks of $G$ and $\hat{K}$, whereas the downstream feedback system is composed of $G$ and $\overline{G}$.
Focusing on the upstream feedback system, we can design $\hat{K}$ using a standard controller design technique for $G$ that is ``isolated" from $\overline{G}$.
More specifically, the output signals $\hat{z}$ and $\hat{w}$ from the upstream feedback system can be directly regulated based on a suitable choice of $\hat{K}$.
On the other hand, $z$ is shown to be the sum of $\hat{z}$ and $\check{z}$, and the latter depends on $\overline{G}$ as
\[
\check{z}=G_{zv}(I-\overline{G}G_{wv})^{-1}\overline{G}\hat{w}.
\]
Therefore, we see that a design criterion of $\hat{K}$ should, in principle, be specified for $\hat{z}$ and $\hat{w}$.
In Section~\ref{sec:2stage}, we will discuss the appropriate design criterion, when multiple retrofit controllers are simultaneously designed and implemented.

\begin{remark}\label{remout}
The output-rectifying retrofit controller shown in Proposition~\ref{thm:out_retro} is essentially identical to that derived in our previous work \cite{ishizaki2018retrofit}.
The novelty of Proposition~\ref{thm:out_retro}, as compared to the existing result, is to clarify that ``all" such output-rectifying retrofit controllers can be expressed as the unique form of \eqref{eq:out_retro}.
This uniqueness is proven by the constrained Youla parameterization in Theorem~\ref{thm:main}.
A generalization of Proposition~\ref{thm:out_retro} to unstable subsystems can be found in Appendix~\ref{app:general}.
The cascade structure shown in Proposition~\ref{co:cl_out} is itself not new; however its derivation provides a frequency-domain analysis analogous to the state-space analysis in the previous work, which can also be conducted without Assumption~\ref{assum:loc_stable}.
\end{remark}

\section{Analysis from Perspective of Multiple Designers}\label{sec:2stage}

With regard to the modular design problem of decentralized controllers mentioned in Section~\ref{subsecprob}, on comparing Fig.~\ref{fig:ent_N} and Fig.~\ref{fig:pre_sys}, we find that it can be regarded as a ``macroscopic" retrofit control problem where the interaction $\bm{L}$ corresponds to the environment for the block-diagonally structured system $\bm{G}$.
Thus, we state the following.

\begin{theorem}\label{probiret}
Let Assumption~\ref{assum:loc_stable} hold for each of all subsystems.
Then, the entire network system in Fig.~\ref{fig:ent_N} is internally stable for any interaction $\bm{L}$ such that the preexisting system $\bm{G}_{\rm pre}$ is internally stable if and only if all subcontrollers $K_1,\ldots K_N$ are retrofit controllers.
\end{theorem}

\begin{IEEEproof}
Regard $\overline{G}$ as $\bm{L}$, and $G_{\rm pre}$ as $\bm{G}_{\rm pre}$ in Definition~\ref{def:retro}.
It is sufficient to prove that $\bm{K}$ is a retrofit controller if and only if all  $K_1,\ldots K_N$ are retrofit controllers.
This is proven because the Youla parameter of $\bm{K}$, i.e.,
\[
\bm{Q}=\bm{K}(I-\bm{G}_{\bm{yu}}\bm{K})^{-1}
\]
is block-diagonal, as both $\bm{K}$ and $\bm{G}_{\bm{yu}}$ are supposed to be block-diagonal.
Therefore, $\bm{G}_{\bm{wu}}\bm{Q}\bm{G}_{\bm{yv}}=0$ holds if and only if $G_{w_i u_i }Q_i G_{y_i v_i}=0$ holds for all $i = 1,\ldots, N$.
\end{IEEEproof}

For the modular design problem of decentralized controllers, Theorem~\ref{probiret} shows that the set of retrofit controllers, i.e., 
\[
\mathcal{M}_{i} = 
\left\{
K_i : G_{w_i u_i }K_i (I-G_{y_i u_i} K_i)^{-1} G_{y_i v_i}=0
\right\},
\]
is only the set of subcontrollers such that the system stability is guaranteed for any interaction $\bm{L}$ such that the preexisting system $\bm{G}_{\rm pre}$ is internally stable.
Next, we focus on the output-rectifying retrofit control, based on the premise that Assumption~\ref{assum:v} holds for each of all subcontroller designers.
In the following, Assumption~\ref{assum:loc_stable} is not necessary because it is not essential to prove the ``if" part of Proposition~\ref{thm:out_retro} and Proposition~\ref{co:cl_out}.
We analyze the structure of the entire feedback system map when simultaneously implementing multiple output-rectifying retrofit controllers.
The following claim is proven by replacing $\overline{G}$ with $\bm{L}$, and $G$ with $\bm{G}$ in Proposition~\ref{co:cl_out}.

\begin{proposition}\label{lementcon}
Let Assumption~\ref{assum:v} hold for each of all subcontroller designers.
Consider that every $K_i$ is an output-rectifying retrofit controller in Proposition~\ref{thm:out_retro}.
Then, the entire map $\bm{T}_{\bm{z}\bm{d}}: \bm{d} \mapsto \bm{z}$ compatible with Fig.~\ref{fig:ent_N} is structured as
\begin{equation}\label{Tzdentret}
\spliteq{
\bm{T}_{\bm{z}\bm{d}}(
\hat{K}_1, & \ldots,\hat{K}_N;\bm{L})
  = 
\sfdiag \bigl(\hat{M}_{z_i d_i}(\hat{K}_i)\bigr)  \\
&+  \bm{G}_{\bm{zv}}
(I-\bm{L}\bm{G}_{\bm{wv}})^{-1} \bm{L}\ \!
\sfdiag \bigl(\hat{M}_{w_i d_i}(\hat{K}_i)\bigr).
}
\end{equation}
\end{proposition}

It is evident that the cascade structure shown in Proposition~\ref{co:cl_out} is also proven for the entire map of Fig.~\ref{fig:ent_N}.
Notably, the terms relevant to $\hat{M}_{z_i d_i}$ and $\hat{M}_{w_i d_i}$ on the right-hand side of \eqref{Tzdentret} have a block-diagonal structure and are decoupled from the term relevant to $\bm{L}$.
This implies that each problem of finding  $\hat{K}_i$ can be decoupled with respect to each subcontroller designer, and these subcontrollers do not affect the initial system stability premised in Assumption~\ref{assentsys}\textbf{(a)}.

For any interaction $\bm L$ such that the preexisting system ${\bm G}_{\rm pre}$ in \eqref{platsys} is internally stable, there exists $\delta \geq 0$ such that 
\begin{equation}\label{exJ}
 \|\bm{G}_{\bm{zv}}(I-\bm{L}\bm{G}_{\bm{wv}})^{-1}\bm{L}\|_{\infty}\leq \delta.
\end{equation}
Using this value of $\delta$, we can derive the following bound of the entire control performance.

\begin{proposition}\label{thm:cl_all}
Let Assumption~\ref{assum:v} hold for each of all subcontroller designers.
Consider that every $K_i$ is an output-rectifying retrofit controller in Proposition~\ref{thm:out_retro}. 
If
\begin{equation}\label{exJi}
\bigl\|\hat{M}_{z_i d_i}(\hat{K}_i)\bigr\|_{\infty} \leq {\alpha_i},\quad
\bigl\|\hat{M}_{w_i d_i}(\hat{K}_i)\bigr\|_{\infty}\leq \beta_i
\end{equation}
for the design of each subcontroller, then
\begin{equation}\label{exbnd}
\|\bm{T}_{\bm{zd}} \|_{\infty}\leq 
\sfmax_{i} \alpha_i + \delta \sfmax_{i} \beta_i
\end{equation}
for the entire map $\bm{T}_{\bm{zd}} :\bm{d}\mapsto \bm{z}$ compatible with Fig.~\ref{fig:ent_N}.
\end{proposition}

\begin{IEEEproof}
Applying the triangular inequality to \eqref{Tzdentret} and using the submultiplicativity of the $\mathcal{H}_{\infty}$-norm, we can easily obtain the bound of \eqref{exbnd}, using \eqref{exJ} and \eqref{exJi}.
\end{IEEEproof}

Proposition~\ref{thm:cl_all} identifies a monotonically increasing function $\gamma$ in the modular design problem of decentralized controllers.
This is shown below.
If either $\|\hat{M}_{z_i d_i}\|_{\infty}$ or $\|\hat{M}_{w_i d_i}\|_{\infty}$ is constrained by a given bound, then the resultant upper bound of $\|\bm{T}_{\bm{zd}} \|_{\infty}$ is found to be monotonically increasing for each objective value.
In particular, if the design criterion for each subcontroller designer is determined as
\begin{equation}\label{objJi}
\spliteq{
J_i\bigl[ \hat{M}_i(\hat{{K}}_i) \bigr] & = 
\bigl\|
\hat{M}_{z_i d_i}(\hat{K}_i)
\bigr\|_{\infty} \\ 
&\textrm{subject to} \quad
\bigl\|\hat{M}_{w_i d_i}(\hat{K}_i)\bigr\|_{\infty}\leq \beta_{i},
}
\end{equation}
where $\beta_i$ denotes a given bound, then
\begin{equation}
\gamma(J_1^{\star},\ldots,J_N^{\star})=\sfmax_{i} J_i^{\star} + \delta \sfmax_{i} \beta_i
\end{equation}
is found to be a bounding function.
Because $\hat{K}_i$ is a stabilizing controller just for $G_{y_i u_i}$, such a subcontroller design problem can be  handled via existing approaches, such as the $\mu$-synthesis or others.
It should be noted that, in principle, any existing controller design method can be applied to determine a possible $\hat{K}_i$.
This offers feasible options to each subcontroller designer when designing their subcontroller depending on the intended applications.
We remark that it may be difficult to determine the actual value of $\delta$ in practice because $\bm L$ is supposed to involve unknown parameters.
The significance of Proposition~4 is to prove that, for any admissible $\bm L$, there exists a bounding function, dependent on $\bm L$, such that it is monotone increasing for the local performance indices $J_1^{\star},\ldots,J_N^{\star}$.
We note that a similar upper bound  in terms of the $\mathcal{H}_2$-norm or others can also be derived from  \eqref{Tzdentret}.

\begin{remark}
Every output-rectifying retrofit controller is ``self-responsible" for its local disturbance attenuation.
In particular, it can be shown that 
$u_i = 0$ for any $d_j \in \mathcal{D}_j$ such that $j\neq i$,
where $\mathcal{D}_j$ denotes the set of all possible disturbance inputs. 
This implies that an output-rectifying retrofit controller implemented to the $i$th subsystem is ``insensitive" to the disturbances injected to any other subsystems.
Conversely, the $i$th output-rectifying retrofit controller works only for its own disturbance $d_i$.
Therefore, if all subcontroller designers adopt the output-rectifying retrofit control, local disturbances occurring in individual subsystems are to be managed on their own responsibility.
\end{remark}


\begin{figure*}[t]
\centering
\includegraphics[width = .95\linewidth]{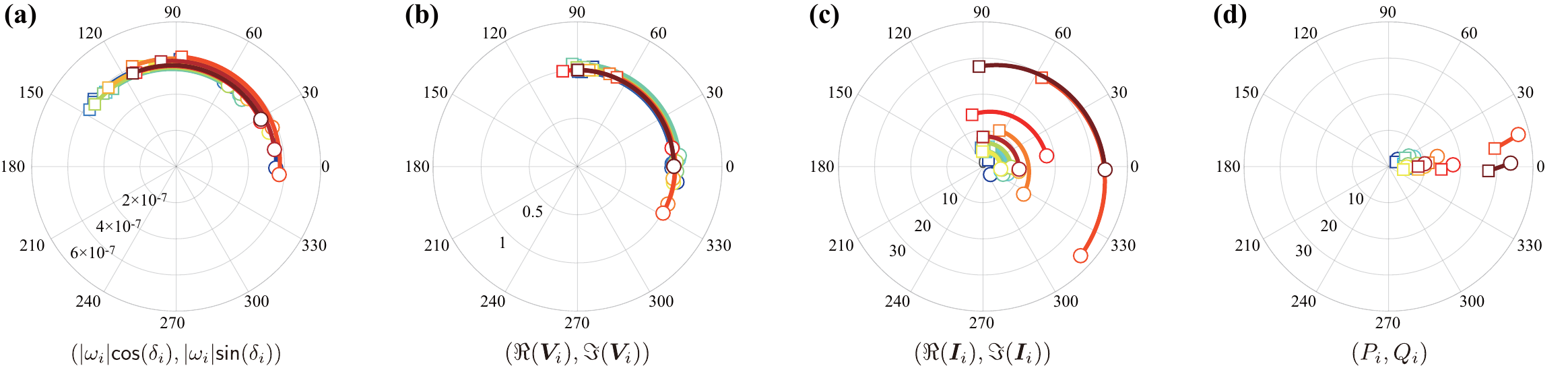}
\caption{
Loci of $(|\omega_i| \sfcos(\delta_i),|\omega_i| \sfsin(\delta_i))$, $(\Re(\bm{V}_i),\Im(\bm{V}_i))$,  $(\Re(\bm{I}_i),\Im(\bm{I}_i))$, and $(P_i,Q_i)$ in response to load variation, where
$\delta_i$ is the rotor angle, 
$\omega_i$ is the frequency deviation, 
$\bm{V}_i$ is the voltage phasor, 
$\bm{I}_i$ is the current phasor,
$P_i$ is the active power, and
$Q_i$ is the reactive power.
Details of these physical variables are presented in Appendix~\ref{apx_pmod}.
The circles and squares correspond to the initial and final load impedances, respectively.
The final load impedances are 30\% lower than the initial load impedances.}
\label{figtphaseplot}
\end{figure*}

\begin{figure}[t]
\centering
\includegraphics[width = .99\linewidth]{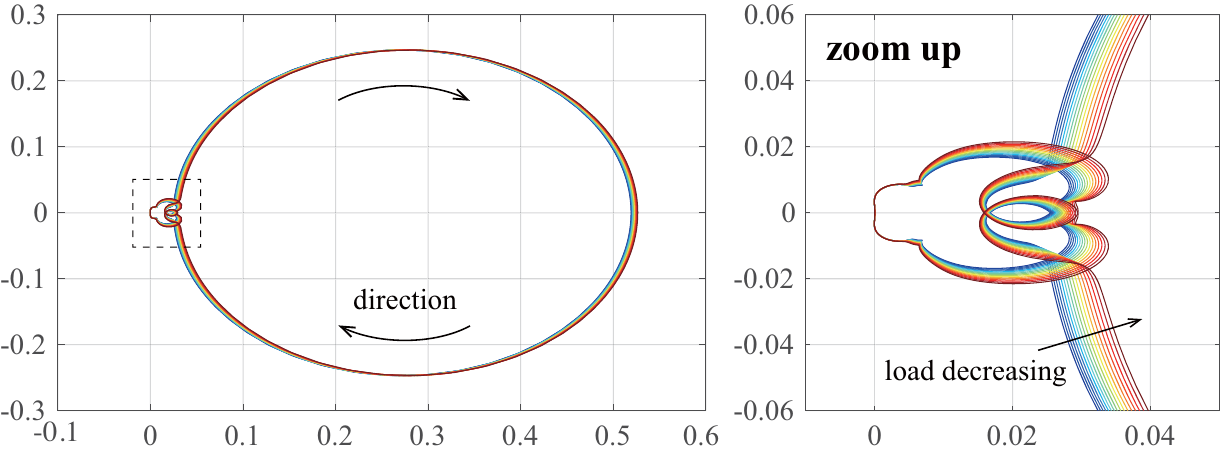}
\caption{Nyquist plots of power systems linearized at operating points.}
\label{fignyquist}
\end{figure}

\section{Numerical Demonstration}\label{secnum}

Considering the example of GFC presented in Section~\ref{secmotex}, we demonstrate the significance of the proposed modular design method of decentralized controllers.
Moreover, we show that the overall control performance can be improved sequentially as the number of implemented retrofit controllers, i.e., PSS upgrade modules, increases.

\subsection{AGC Based on Broadcast-Type PI Control}\label{sec_nAGC}

First, we demonstrate AGC implemented as a broadcast-type PI control in \eqref{gPI}.
Using the nonlinear model in \eqref{gendyn}--\eqref{powbal}, we observe the transition of the system variables in response to load variation.
We consider varying the load impedance set $\{Z_i\}_{i \in \mathcal{I}_{\rm L}}$ in (\ref{dynload}) at a linear rate such that all impedances decrease by 10\% in an hour.
For an interval of three hours, we plot the resultant loci of the generator variables in Fig.~\ref{figtphaseplot}.
From Fig.~\ref{figtphaseplot}\textbf{(a)}, we see that the frequency deviations of all generators are almost synchronized, and they are sufficiently small, while the rotor angles gradually vary in response to the load variation.
Fig.~\ref{figtphaseplot}\textbf{(b)} shows that the voltage amplitudes at all generator buses remain almost identical and non-oscillatory.
This implies that the AGC maintains the entire system in a quasi-steady state.
Figs.~\ref{figtphaseplot}\textbf{(c)} and \textbf{(d)} show that, in response to load variation, each generator reduces its active power output by decreasing its current amplitude.

It should be noted that the AGC attains stability at all operating points of interest.
This can be verified based on a passivity-short property of the power system.
To observe this, we plot the Nyquist plots of the linearized version of \eqref{gendyn}--\eqref{powbal} at 16 operating points obtained during load variation.
Their input and output ports are chosen as being compatible with the AGC in \eqref{gPI}, where  the broadcast input $\overline{U}$ is applied as a scalar input to the entire network system, and the average frequency deviation $\overline{\omega}$ is measured as a scalar output from the system.
Therefore, considering the ports of $\overline{U}$ and $\overline{\omega}$ as the input and output ports, respectively, we obtain a single-input single-output system associated with each operating point.
From Fig.~\ref{fignyquist}, where the blue and red lines correspond to the initial and final load impedances, respectively, we see that all the transfer functions that we have obtained are positive real at almost all frequencies.
A similar trend can be observed also for load models other than the constant impedance model.
Thus, system stability can be attained and explained for a majority of PI controllers with moderate gains.

\subsection{GFC Based on Retrofit Control}

\subsubsection{Design of PSS Upgrade Modules}

We design the PSS upgrade modules based on output-rectifying retrofit control, supposing that both $V_i$ and $U_i$ in \eqref{gendyn} are measurable as the interaction input signals.
Note that $V_i$ can be measured via a phase measurement unit (PMU) attached to the $i$th generator bus, in practice.
Moreover, $U_i$ is also available as a reference signal.
In addition, we suppose that the generator state $x_i$ is measurable as a measurement output.
Although the direct measurement of a rotor angle in an absolute frame may not be easy, we can locally estimate it based on methods such as Kalman filtering, provided that the phase of the corresponding generator bus voltage is measured via PMU.

\begin{figure*}[t]
\centering
\includegraphics[width = .95\linewidth]{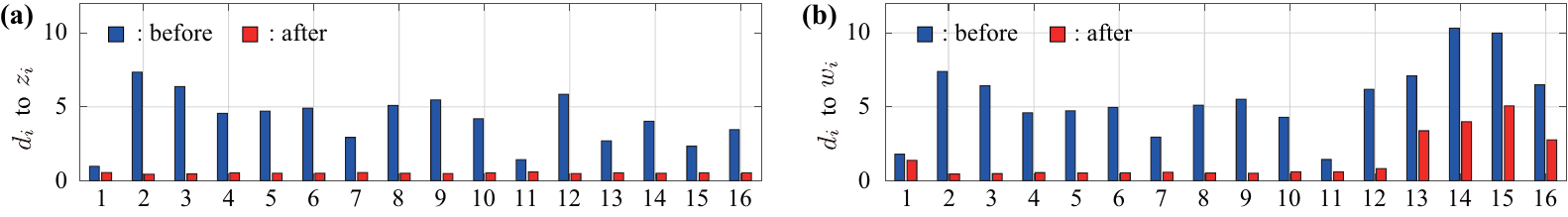}
\caption{
Resultant $\mathcal{L}_2$-gains.
\textbf{(a)} From $d_i$ to $z_i$.
\textbf{(b)} From $d_i$ to $w_i$.
The blue and red bars quantify the $\mathcal{L}_2$-gains before and after controller design,  respectively.
The numbers on the horizontal axis represent the labels of generator buses.}
\label{fignorms}
\end{figure*}

For the design of the PSS upgrade modules, each PSS designer aims at determining a suitable $\hat{K}_i$ in Proposition~\ref{thm:cl_all}, such that it stabilizes
\[
G_{y_i u_i}(s)=(sI-A_i^{\star})^{-1}B_i^{\star},
\]
where $A_i^{\star}$ and $B_i^{\star}$ are given as in \eqref{lingendyn}.
As a design criterion, the gains of \eqref{exJi} for the isolated subsystems
\[
G_{z_i d_i}(s)=S_i(sI-A_i^{\star})^{-1},\quad
G_{w_i d_i}(s)=
\mat{
S_i\\
\mathit{\Gamma}_i^{\star}
}
(sI-A_i^{\star})^{-1}
\]
are considered, where the disturbance input port matrix is chosen as the identity matrix because different faults stimulate each generator state almost randomly.
In addition, the output ports of both $\omega_i$ and $I_i$ are identified as the interaction output ports because these are the interaction signals outflowing to the broadcast-type PI controller in \eqref{gPI} and the transmission network, respectively.
In this setting, a set of stabilizing controllers is found using linear quadratic regulator (LQR) design.
The resultant $\mathcal{L}_2$-gains before and after the controller design, i.e., the $\mathcal{H}_{\infty}$-norms of the isolated subsystems with and without the stabilizing controllers, are plotted as the blue and red bars, respectively, in Fig.~\ref{fignorms}.
This indicates that each $\hat{K}_i$ is designed suitably.
We remark that the GFC is formulated based on the linearization around an operating point, implying that each PSS upgrade module aims at restoring the corresponding generator state to its operating point before the occurrence of a ground fault.

\begin{figure*}[t]
\centering
\includegraphics[width = .95\linewidth]{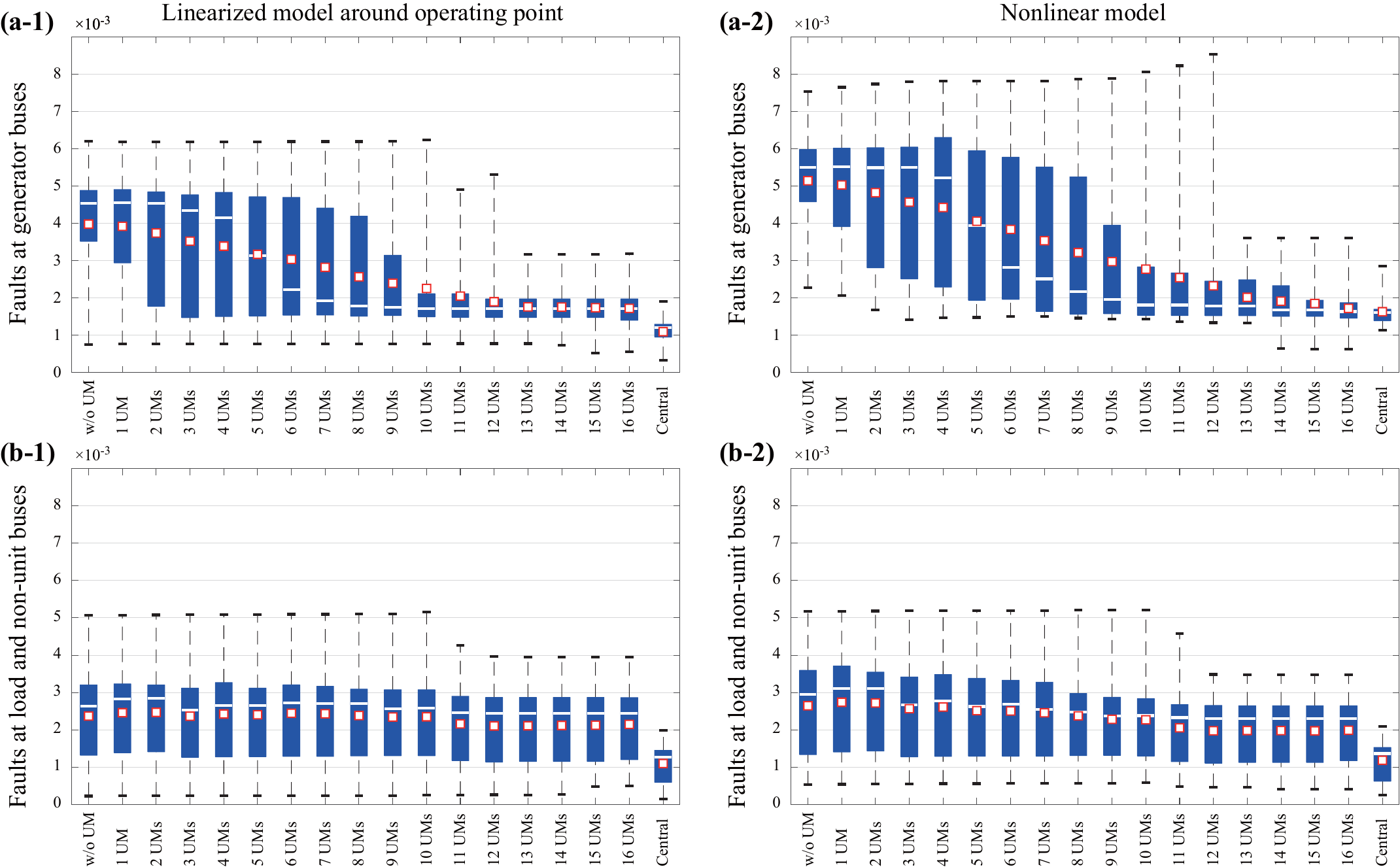}
\caption{Box plots of the magnitude of frequency deviation with respect to the number of implemented PSS upgrade modules (UMs). The upper and lower rows correspond to the faults on the generator buses and those on the load and non-unit buses, respectively.
The left and right columns correspond to the cases of linearized and nonlinear models, respectively.}
\label{figboxplots}
\end{figure*}

\begin{figure*}[t]
\centering
\includegraphics[width = .99\linewidth]{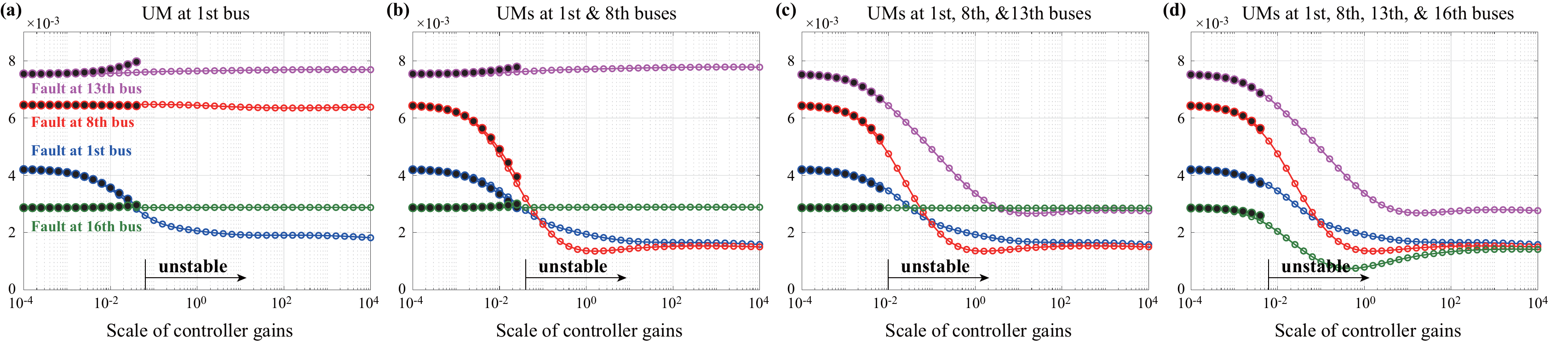}
\caption{
The values of $\|\omega^{(1)}\|_{\mathcal{L}_2}$, $\|\omega^{(8)}\|_{\mathcal{L}_2}$, $\|\omega^{(13)}\|_{\mathcal{L}_2}$, and $\|\omega^{(16)}\|_{\mathcal{L}_2}$ versus the scale of controller gains.
The lines with the black circles correspond to the simple decentralized control, while those with the colored circles correspond to the proposed retrofit control.
A PSS upgrade module is denoted as ``UM."
}
\label{figcompLQR}
\end{figure*}

\vspace{1mm}
\subsubsection{Results}

Denote the frequency deviation of all generators as $\omega := (\omega_1, \ldots , \omega_{16})$.
To represent its dependence on each fault, we denote the resultant $\omega$ in response to the $i$th bus fault as $\omega^{(i)}$.
Using the $\mathcal{L}_2$-norm as a performance measure, we aim at appropriately reducing the magnitude of $\|\omega^{(i)}\|_{\mathcal{L}_2}$ for all $i = 1,\ldots, 68$.
To visualize the significance of PSS upgrade based on retrofit control, we draw the box plots of the datasets $\{\|\omega^{(i)}\|_{\mathcal{L}_2}\}_{i = 1}^{16}$ and $\{\|\omega^{(i)}\|_{\mathcal{L}_2}\}_{i = 17}^{68}$,  corresponding to the faults on the generator buses and the load and non-unit buses, respectively.

We show the results when using the linearized model around the operating point indicated by the circles in Fig.~\ref{figtphaseplot}.
When a PSS upgrade module is not implemented, the resultant box plots are depicted as the first columns in Figs.~\ref{figboxplots}\textbf{(a-1)} and \textbf{(b-1)}, where we denote the minimum and maximum by the top and bottom black bars, the median by the white bar, and the average by the square mark.
It can be seen that the frequency deviations arising due to generator bus faults are generally greater than those arising due to load and non-unit bus faults.
This is normal because the generators are directly connected to the generator buses, and not to the other buses.

To simulate the gradual penetration of the PSS upgrade, we sequentially implement the designed PSS upgrade modules from the 1st to the 16th generators.
The resultant box plots with respect to the number of implemented PSS upgrade modules are depicted in the columns from the second in Figs.~\ref{figboxplots}\textbf{(a-1)} and \textbf{(b-1)}.
From Fig.~\ref{figboxplots}\textbf{(a-1)}, we see that the frequency deviations due to the generator bus faults are gradually reduced as the number of implemented modules increases.
On the other hand, from Fig.~\ref{figboxplots}\textbf{(b-1)}, we see that, even though the PSS upgrade is not very sensitive to load and non-unit bus faults, the maximum values are reduced when the PSS upgrade modules penetrate significantly.
Additionally, similar box plots were obtained for other operating points.

For reference, we consider designing a ``centralized" LQR assuming the availability of the entire power system model.
The resultant box plots are depicted in the last columns of Figs.~\ref{figboxplots}\textbf{(a-1)} and \textbf{(b-1)}; they indicate that the performance of the upgraded PSSs approximates the best achievable performance of the centralized LQR.
For further comparison, we calculate the same box plots for the nonlinear power system model, while implementing
a nonlinear extension of the output rectifier reported in \cite{ishizaki2018retrofit,sadamoto2018retrofit}, where the nonlinearity neglected in controller design is regarded as a part of the environments.
From the resultant box plots in Figs.~\ref{figboxplots}\textbf{(a-2)} and \textbf{(b-2)}, we see that the performance of GFC for the nonlinear model is almost comparable with that for the linearized model, especially when sufficient upgraded PSSs are implemented.
This implies that the upgraded PSSs successfully confine state deviations to a domain close to the operating point, wherein nonlinearity is almost negligible.

Finally, we demonstrate that simple decentralized control can easily destabilize the power system.
In particular, we consider a PSS upgrade module given as $u_i = \hat{K}_i x_i$, where a stabilizing controller $\hat{K}_i$ for the isolated subsystem $G_{y_i u_i}$ is just implemented without the output rectifier.
Implementing such a simple module to the first generator, we plot the resultant values of $\|\omega^{(i)}\|_{\mathcal{L}_2}$ for $i=1$, 8, 13, 16 in Fig.~\ref{figcompLQR}\textbf{(a)}, where the scale of controller gains is varied.
In this figure, the results obtained by the simple module are shown by the lines with the black circles, while the results obtained by the proposed module involving the abovementioned nonlinear output rectifier are also shown by the lines with the colored circles for comparison.
Note that the lines with the black circles break off in the middle because of instability.
In contrast, the proposed module can improve the control performance for the first bus fault, as the controller gain increases.
Figs.~\ref{figcompLQR}\textbf{(b)}--\textbf{(d)} show the results when multiple modules are implemented.
Each value of $\|\omega^{(i)}\|_{\mathcal{L}_2}$ is reduced by the proposed module implemented to the corresponding generator, while instability is induced in every case where the simple modules are implemented. 
These results clearly illustrate the significance of the PSS upgrade based on the retrofit control.

\section{Concluding Remarks}

In this paper, a modular design method of decentralized controllers has been developed for linear dynamical network systems.
Modular design or modularity-in-design is a widely accepted concept in system design; it simplifies complex large-scale system design, enables parallel work by multiple independent entities, and enables flexible future modifications of modules.
As illustrated in the example of power systems control, a manageable complexity of system modeling and controller design is obtained by enabling parallel work via multiple subcontroller designers.
Moreover, the flexibility in designing and implementing respective controllers is ensured, because each designer can individually add, remove, and modify their controller without considering the actions of other designers.

In this paper, we assume  that no information regarding the environments is unavailable during the design of retrofit controllers.
Such a strict information constraint may restrict the performance improvement depending on the intended applications.
Regarding this concern, our recent work \cite{ishizaki2019retrofit} has proposed a sophisticated technique to utilize ``partial" information of environments.
It is shown that an approximate environment model can be used to determine a better stabilizing controller during controller design.
The integration of this technique further strengthens the efficacy of retrofit control.
Furthermore, we do not discuss a situation where not only subcontrollers but also subsystems themselves are  modified.
To discuss this situation, we require a more detailed analysis of the interaction among subsystems, considering the robustness to uncertainty or flexibility of subsystem variation, for which relaxing the constraint on the Youla parameter in a robust control setting would be noteworthy. 
A related result based on a small gain condition can be found in \cite{tan1990decentralized}.
In addition, depending on the applications, subsystem partition may not be well-defined and may also affect the control performance of the resultant feedback system.
For the decomposition of dynamical networks, methods such as a nested decomposition method \cite{sezer1986nested} and a computationally efficient community detection method \cite{tang2018optimal,tang2018network} would be promising, provided that the global information regarding the entire network model is available, at least partially.
Discussions on these open issues are possible directions of our future research.


%

\appendices

\section{Flux-Decay Model of Generators}\label{apx_pmod}

For reference, we provide a brief review of generator  models.
A comprehensive list of of power system component models can be found in \cite{sadamoto2019dynamic}.
In this appendix, phasors are denoted by the bold face symbols, such as $\bm{I}_i$.
Using this notation, the current $I_i$ and voltage $V_i$ in \eqref{gendyn} are given as 
\[
I_i =\sfcol(\Re (\bm{I}_i),\Im (\bm{I}_i)),\quad 
V_i = \sfcol(\Re (\bm{V}_i),\Im (\bm{V}_i)),
\]
respectively.
For simplicity, we neglect the generator bus label ``$i$" in the following.

\begin{subequations}\label{spgen}
A standard generator model, known as the one-axis model or flux-decay model, is reviewed.
The electromechanical swing dynamics with electromagnetic dynamics is given as
\begin{equation}\label{spgensd}
\spliteq{
\dot{\delta}&= \omega_0  \omega\\
M   \dot{\omega}&= 
 - D  \omega +U - P \\
\tau_{\rm d} \dot{E} &= \textstyle
 -\frac{X_{\rm d}}{X_{\rm d}^{\prime}}E
+\left(
\frac{X_{\rm d}}{X_{\rm d}^{\prime}}-1
\right)
|\bm{V}| \sfcos (\delta - \angle \bm{V} ) 
+ V_{\rm{field}}
}
\end{equation}
where $\delta$ denotes the rotor angle relative to the frame rotating at the standard frequency $\omega_0$, 
$ \omega$ denotes the frequency deviation relative to $\omega_0$,
$E$ denotes the q-axis voltage behind the transient reactance $X_{\rm d}$, 
$X_{\rm d}^{\prime}$ denotes the d-axis transient reactance, 
$M$ denotes the inertia constant, 
$D$ denotes the damping coefficient, 
$\tau_{\rm d} $ denotes the d-axis transient open-circuit time constant, 
$U$ denotes the reference input for mechanical power regulation,
$P$ denotes the active power,
$\bm{V}$ denotes the voltage phasor of the bus connected with the generator, and
$V_{\rm{field}}$ denotes the field voltage.
The active power and reactive power are respectively written as
\[
\spliteq{
P &= \textstyle \frac{|\bm{V}|E}{X_{\rm d}^{\prime}} \sfsin(\delta -  \angle \bm{V})
-  \frac{|\bm{V}|^2}{2}  
\left( \frac{1}{X_{\rm d}^{\prime}}  -  \frac{1}{X_{\rm q}} \right)
\sfsin( 2\delta - 2\angle \bm{V})\\
Q &= \textstyle \frac{|\bm{V}|E}{X_{\rm d}^{\prime}} \sfcos (\delta - \angle \bm{V})
-|\bm{V}|^2 \left( \frac{\sfcos^2 (\delta - \angle \bm{V}) }{X_{\rm d}^{\prime}} 
+ \frac{\sfsin^2 (\delta - \angle \bm{V})}{X_{\rm q}} \right).
}
\]
The excitation system with AVR is modeled as
\begin{equation}
\tau_{\rm e} \dot{V}_{\rm{field}} =
-V_{\rm{field}} + V_{\rm{field}}^{\star}
+ 
K_{\rm AVR}
\bigl(
|\bm{V}| - V^{\star} + u_{\rm PSS} +u
\bigr)
\end{equation}
where $\tau_{\rm e}$ denotes the exciter time constant, 
$V_{\rm{field}}^{\star}$ denotes the operating point of the field voltage,
$K_{\rm AVR}$ denotes the AVR gain,
$V^{\star}$ denotes the operating point of the bus voltage magnitude, 
$u_{\rm PSS}$ denotes the reference input from PSS, and 
$u$ denotes the additional reference input to AVR.
A standard PSS is considered as a three-dimensional controller that consists of a two-stage lead-lag compensator and a high-pass washout filter.
Because the specific form of PSS is not relevant to the discussion in this paper, we simply write it as
\begin{equation}\label{spgenpss}
u_{\rm PSS} = \mathcal{K}_{\rm PSS} ( \omega)
\end{equation}
where $\mathcal{K}_{\rm PSS}$ denotes a linear map such that $\mathcal{K}_{\rm PSS}(0)=0$.
The specific form of $\mathcal{K}_{\rm PSS}$ and standard values of the constant parameters can be found in \cite{sadamoto2019dynamic}.
\end{subequations}
Finally, the current phasor flowing from the generator to the generator bus is given as 
\begin{equation}\label{spgencr}
\textstyle
\bm{I}=
e^{j \delta}
\left(
\frac{ |{\bm V}|\sfsin(\delta -  \angle \bm{V}) }{X_{\rm q}} +
j \frac{  |{\bm V}|\sfcos(\delta -  \angle \bm{V}) -E }{ X_{\rm d}^{\prime} }
\right).
\end{equation}
It can be verified that  ${\bm V} \overline{{\bm I}} = P +jQ$, where the over-line stands for the complex conjugate.
The function $f_i$  in \eqref{gendyn} can be identified from \eqref{spgen}, and $g_i$  can be identified from \eqref{spgencr}.

\section{Proof of Theorem~\ref{thm:main}}\label{sec:pfthm}

\begin{figure}[t]
\centering
\includegraphics[width = .55\linewidth]{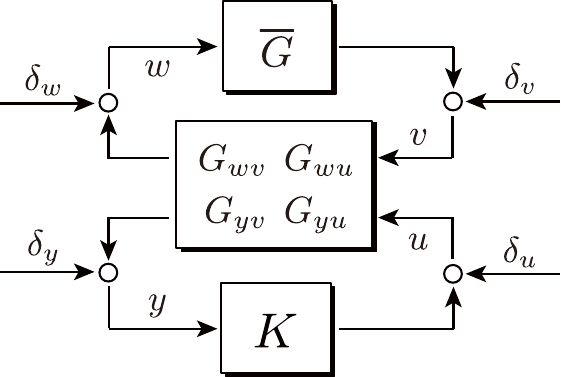}
\caption{Equivalent block diagram for stability analysis.}
\label{fig:sta_ana}
\end{figure}

First, we prove the sufficiency, i.e., if $K$ written by \eqref{eq:youK} satisfies \eqref{eq:gencond_sta}, then $K$ is a retrofit controller.
The internal stability of Fig.~\ref{fig:pre_sys} is equivalent to that of Fig.~\ref{fig:sta_ana} owing to \cite[Lemma~12.2]{zhou1995robust}.
Its internal stability is proven if the sixteen transfer matrices from $(\delta_u, \delta_y, \delta_{v}, \delta_{w})$ to $(u, y, v, w)$ all belong to $\mathcal{RH}_{\infty}$.

When $K=0$, Fig.~\ref{fig:sta_ana} is internally stable because $G_{\rm pre}$ in \eqref{presys} is assumed to be internally stable.
From the system of the algebraic equations
\begin{equation}\label{algeq}
\simode{
w &= G_{wv}v + G_{wu} u + \delta_w \\
v &= \overline{G}w + \delta_v\\
y &= G_{yv}v + G_{yu} u + \delta_y\\
u &= Ky + \delta_u
}
\end{equation}
with $K=0$, we have
\[
\spliteq{
y= 
\delta_y  &+
\bigl\{
\underbrace{
G_{yu} + G_{yv}\overline{G}(I-G_{wv}\overline{G})^{-1} G_{wu} 
}_{T_{yu}}
\bigr\} \delta_u 
\\
&+
\underbrace{
G_{yv} (I-\overline{G}G_{wv})^{-1}
}_{T_{yv}}
 \delta_v +
\underbrace{
G_{yv} \overline{G}(I-G_{wv}\overline{G})^{-1}
}_{T_{yw}}
\delta_w,
}
\]
where all $T_{yu}$, $T_{yv}$, and $T_{yw}$ belong to $\mathcal{RH}_{\infty}$.
In the same manner, if $K\neq 0$, we have
\begin{equation}\label{sigy}
y = \delta_y  + T_{yu} u  + T_{yv} \delta_v + T_{yw} \delta_w,
\end{equation}
where $u$ is a function of $(\delta_u, \delta_y, \delta_{v}, \delta_{w})$.
Note that $T_{yu}$ can be written as
\[
T_{yu} = G_{yu} + 
\underbrace{
G_{yv} \overline{Q} G_{wu}
}_{X(\overline{Q})}.
\]
Substituting them into the equation of $u$ in \eqref{algeq}, we have
\[
(I- K G_{yu}) u =  K X(\overline{Q})u + K 
\left\{
\delta_y   + T_{yv} \delta_v + T_{yw} \delta_w
\right\}
+ \delta_u.
\]
Multiplying it by $(I- K G_{yu})^{-1}$ from the left side and solving it with respect to $u$, we see that
\begin{equation}\label{sigu}
\spliteq{
u = &\bigl(I- QX(\overline{Q})\bigr)^{-1}  \\
& \times
\left\{
Q
(
\delta_y   + T_{yv} \delta_v + T_{yw} \delta_w
)
+ (I+Q G_{yu})\delta_u
\right\}
}
\end{equation}
where we have used the relations
\[
Q=(I-KG_{yu})^{-1}K
,\quad
(I- K G_{yu})^{-1} = I+Q G_{yu}.
\]
Furthermore, we see that
\[
\bigl(I- QX(\overline{Q})\bigr) \bigl(I+ QX(\overline{Q})\bigr)
=
I - \left( QX(\overline{Q}) \right)^2
\]
where the square term is zero if  \eqref{eq:gencond_sta} holds.
This means that
\[
\bigl(I- QX(\overline{Q})\bigr)^{-1} = I+ QX(\overline{Q}), 
\]
which belongs to $\mathcal{RH}_{\infty}$.
Hence, all the transfer matrices of $u$ in \eqref{sigu} as well as those of $y$ in \eqref{sigy} belong to $\mathcal{RH}_{\infty}$.
In a similar manner, we can verify the stability of all the transfer matrices of $v$ and $w$.
This proves the sufficiency.

Next, we show the necessity, i.e., if $K$ is a retrofit controller, then $K$ is written by \eqref{eq:youK} and satisfies \eqref{eq:gencond_sta}.
For $K$ to be a retrofit controller, Fig.~\ref{fig:sta_ana} must be internally stable even for the choice of $\overline{Q} = 0$, which implies $X(\overline{Q}) =0$.
In this case, $u= Q\delta_y$ if all $\delta_u$, $\delta_v$, and $\delta_w$ are chosen to be zero.
Therefore, $Q$ must be in $\mathcal{RH}_{\infty}$, meaning that $K$ is necessarily a stabilizing controller for $G_{yu}$.

What remains to show is the necessity of \eqref{eq:gencond_sta}.
To prove the claim via contradiction, let us suppose that \eqref{eq:gencond_sta} does not hold.
Then, as shown in the proof of the small-gain theorem \cite[Theorem~9.1]{zhou1995robust}, there exists some $\overline{Q} \in \mathcal{RH}_{\infty}$ such that
\[
\sfdet \bigl(
I-
G_{wu}(j \omega_0) 
Q(j \omega_0) 
G_{yv}(j \omega_0) 
\overline{Q}(j \omega_0) 
\bigr) = 0
\]
for some $\omega_0 \in \mathbb{R}\cup\{\infty\}$.
Note that
\[
\sfdet \bigl(I- QX(\overline{Q})\bigr)
= 
\sfdet \bigl(
I-
G_{wu}
Q
G_{yv}
\overline{Q}
\bigr)
\]
because of Sylvester's determinant identity \cite[Fact 2.14.13]{bernstein2009matrix}.
This implies the instability of the inverse term in \eqref{sigu}, which contradicts the internal stability of Fig.~\ref{fig:sta_ana}.

\section{Generalizations to Unstable Subsystems}\label{app:general}

\begin{figure}[t]
\centering
\includegraphics[width = .45\linewidth]{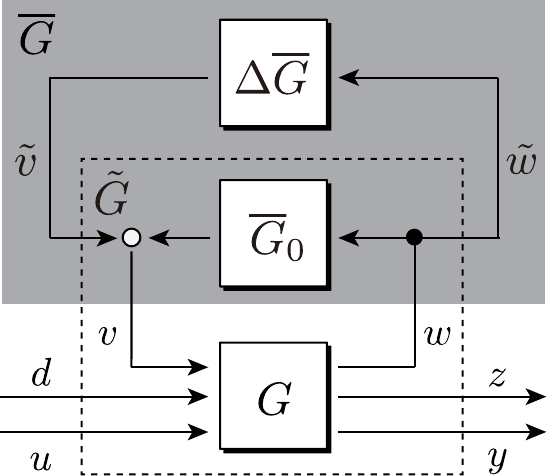}
\caption{Decomposition of environment.}
\label{fig:newpre}
\end{figure}

We first generalize Theorem~\ref{thm:main} as removing Assumption~\ref{assum:loc_stable}.
To this end, we consider decomposing $\overline{G}$ as
\begin{equation}\label{decenv}
\overline{G} = 
\overline{G}_0 + \Delta \overline{G}
\end{equation}
where $\overline{G}_0$ is some transfer matrix such that the feedback system $\mathcal{F}(\overline{G}_0,G)$ is internally stable.
This decomposition is illustrated in Fig.~\ref{fig:newpre}.
Furthermore, we define 
\begin{equation}\label{tilG}
\tilde{G} :=
\mat{
\tilde{G}_{wv} & \tilde{G}_{wd} & \tilde{G}_{wu} \\
\tilde{G}_{zv} & \tilde{G}_{zd} & \tilde{G}_{zu} \\ 
\tilde{G}_{yv} & \tilde{G}_{yd} & \tilde{G}_{yu} \\
}
\end{equation}
as the transfer matrix from $(\tilde{v},d,u)$ to $(\tilde{w},z,y)$ in Fig.~\ref{fig:newpre} when the block of $\Delta \overline{G}$ is removed.
For example
\begin{equation}\label{Gyutil}
\tilde{G}_{yu} := G_{yu} + G_{yv} (I- \overline{G}_0 G_{wv})^{-1} \overline{G}_0 G_{wu}.
\end{equation}
The others are defined similarly.
In this system representation, $\tilde{G}$ can be regarded as a new stable subsystem, while $\Delta \overline{G}$ as a new environment such that $\mathcal{F}(\Delta \overline{G},\tilde{G})$, being identical to $G_{\rm pre}$ in \eqref{presys}, is internally stable.
In this notation, most of the arguments in Sections~\ref{sec:retro} and \ref{sec:2stage} can apply in the same way.

In particular, the proof of Theorem~\ref{thm:main} in Appendix~\ref{sec:pfthm} can be modified as follows.
Substituting \eqref{decenv} and
\[
v = \tilde{v} + \overline{G}_0 w
,\quad
w = \tilde{w}
\]
into \eqref{algeq}, we have
\begin{equation}\label{algeqtil}
\simode{
\tilde{w} &= \tilde{G}_{wv}\tilde{v} + \tilde{G}_{wu} u + \tilde{\delta}_w \\
\tilde{v} &= \Delta\overline{G} \tilde{w} + \delta_v\\
y &= \tilde{G}_{yv} \tilde{v} + \tilde{G}_{yu} u + \tilde{\delta}_y\\
u &= Ky + \delta_u
}
\end{equation}
where $\tilde{\delta}_w$ and $\tilde{\delta}_y$ are defined as
\[
\spliteq{
\tilde{\delta}_w &:= (I- \overline{G}_0 G_{wv})^{-1} \delta_w,
\\
\tilde{\delta}_y &:= \delta_y + G_{yv}(I- \overline{G}_0G_{wv})^{-1}\overline{G}_0 \delta_w.
}
\]
Note that \eqref{algeq} and \eqref{algeqtil} have the same form, and the transfer matrix from $(\delta_w,\delta_y)$ to $(\tilde{\delta}_w,\tilde{\delta}_y)$ is stable.
Therefore, the following generalization is obtained.

\begin{theorem}\label{thm:maing}
For a stable subsystem $\tilde{G}$ in \eqref{tilG}, consider that the Youla parameterization of $K$ is given as
\[
K = (I+\tilde{Q}\tilde{G}_{yu})^{-1}\tilde{Q},\quad \tilde{Q}\in \mathcal{RH}_{\infty}
\]
where $\tilde{Q}$ denotes its Youla parameter.
Then, $K$ is a retrofit controller if and only if
\[
 \tilde{G}_{wu}\tilde{Q}\tilde{G}_{yv}=0.
\]
\end{theorem}

A notable difference from Theorem~\ref{thm:main} is that the condition of $K$ is written in terms of $\tilde{G}_{yu}$, $\tilde{G}_{wu}$, and $\tilde{G}_{yv}$, instead of the original ones, implying that we need to find some $\overline{G}_0$ that stabilizes $G$.
It should be noted that such $\overline{G}_0$ can be found without any information regarding the environment $\overline{G}$.
This is because every stabilizing controller for $G$ can be a possible choice of $\overline{G}_0$.
We can apply any standard method to find it.

Next, we consider generalizing Proposition~\ref{thm:out_retro}.
To this end, we replace Definition~\ref{def:out}  and  Assumption~\ref{assum:v} as follows.

\begin{definition}\label{def:outg}
For a stable subsystem $\tilde{G}$ in \eqref{tilG}, $K$ is said to be an \emph{output-rectifying retrofit controller} if
\[
\tilde{Q}\tilde{G}_{yv} = 0
\]
where $\tilde{Q}$ denotes the Youla parameter of $K$ in Theorem~\ref{thm:maing}.
\end{definition}

\begin{assumption}\label{assum:vg}
The interaction signal $\tilde{v}$ is measurable in addition to the measurement output $y$.
\end{assumption}

Then, it is clear that a generalized version of Proposition~\ref{thm:out_retro} can be obtained as follows.

\begin{proposition}\label{thm:out_retrog}
Let Assumption~\ref{assum:vg} hold for a stable subsystem $\tilde{G}$ in \eqref{tilG}.
Then, $K$ is an output-rectifying retrofit controller if and only if 
\[
K = \hat{K}\tilde{R}
\]
where $\hat{K}$ is a stabilizing controller for $\tilde{G}_{yu}$, and
\[
\tilde{R} := \mat{
I & -\tilde{G}_{yv}
}
\]
\end{proposition}

We should discuss the reasonability of Assumption~\ref{assum:vg}.
In fact, the interaction input signal $\tilde{v}$ in Fig.~\ref{fig:newpre} may not be directly measurable, but it can be simply produced as
\[
\tilde{v} = v - \overline{G}_0 w
\]
provided that both $v$ and $w$ are measurable, and the system model of $\overline{G}_0$ is available.
Note that not only $v$ but also $w$ is measurable in many practical situations.
For the power system example in Section~\ref{secnum}, $w$ represents the current flowing from a generator to its bus.
In summary, the resultant output-rectifying retrofit controller $K$ is found to be
\[
u = \hat{K} \left\{y - \tilde{G}_{yv} (v - \overline{G}_0 w) \right\},
\]
which can be designed and implemented in a modular fashion.


\section*{Acknowledgment}
This research was supported by TEPCO Memorial Foundation, and JST-Mirai Program 18077648.

\ifCLASSOPTIONcaptionsoff
  \newpage
\fi



%



\bibliographystyle{IEEEtran}
\bibliography{reference,reference_CREST}

%




\end{document}